\documentclass[12pt]{iopart}
\usepackage{iopams}
\usepackage{graphicx}
\usepackage{subfigure}
\DeclareGraphicsExtensions{.eps,.pdf,.ps}
\begin{document}
\title{Dependence on state preparation of noise induced effects in multiqubit systems }
\author{Athanasios C. Tzemos,  Demetris P.K. Ghikas\footnote{Corresponding Author, Tel: +302610-997460, FAX: +302610-997617}}
\address{$�$ Department of Physics, University of Patras, Patras 26500, Greece}
\ead{tzemos@upatras.gr, ghikas@physics.upatras.gr}
\begin{abstract}
The perturbation of multiqubit systems by an external noise can induce various effects like decoherence, stochastic resonance and anti-resonance, and noise-shielding. We investigate how the appearance of these effects on disentanglement time depends on the initial preparation of the systems. We present results for $2$, $3$ and $4$-qubit chains in various arrangements and observe a clear de\-pen\-dence on the combination of initial geometry of the state space and the placement of noise. Finally, we see that temperature can play a constructive role for the control of these noise induced effects.
\end{abstract}
\submitto{\JPA}
\maketitle
\section{Introduction}
Multiqubit systems are the building blocks for the architecture of quantum information storing and processing \cite{Nielsen}. Their efficiency depends strongly on the possibility to retain quantum coherence. But these quantum systems are physically embedded and constantly interacting with their local quantum and classical environments which affect coherence \cite{Schlosshauer}. Thus being fundamentally open quantum systems, they dissipate and decohere with time constants that depend on their parameters and coupling constants and on the strength of perturbation by external agents. There have been many efforts to increase the decoherence time, either passively by increasing isolation or engineering the existence of decoherence free subspaces \cite{Lidar1} or actively by affecting the dynamics with external controls \cite{Lidar2,Knill,Franco}.

As it has turned out, a counterintuitive and unexpected external perturbation with a positive role is the action by a noise source, quantum or classical. Normally one expects that noise would only increase decoherence in a monotonous way, namely increased destruction of coherence with higher levels of noise. But it has been found that noise may be used to isolate a quantum system, the noise-shielding effect (N.S.), and may influence positively or negatively the decoherence time in a monotonous or non-monotonous way. These are the quantum Zeno effects and the stochastic resonance (S.R.) or stochastic anti-resonance (S.AR.) effects \cite{Misra,Zaffino,Francica,Ando,Rivas,Wellens,Ghikas}. But it appears that the manifestation of these effects depends, apart from the detailed dynamical setup, on the way the system has been prepared.

From our preliminary investigation of noise effects on two-qubit systems  \cite{Ghikas}, it became evident that certain classes of initial states are more sensitive than others. Considering this not to be accidental, we extended our study of the state preparation dependence of the noisy perturbations, to $3$ and $4$-qubit Heisenberg XY chains. We have found a clear correlation between the way a quantum system of qubits reacts to an external classical noise and the geometry of the initial states. As a tool we used a Master equation for modeling the dissipative and decoherence processes, and applied an external classical gaussian white noise as a stochastic control. Our investigation is based on the decoherence properties of a 2-qubit subsystem. In the case of two qubits we have only the bosonic environment and the external classical noise. In the cases of $3$ and $4$-qubit systems we consider the extra qubits, after they are traced out,  as a form of local fermionic environment. We quantify the entanglement with the concurrence of the 2-qubit subsystem. We study numerically the dependence of disentanglement time (or entanglement sudden death (E.S.D.)) \cite{Yu1,Yu2,Yu3,Yu4,Yu5,Yu6,Cunha} on the strength of the applied noise for zero and nonzero temperature and for various initial states. We have observed the  effects of increase of decoherence, stochastic resonance, stochastic anti-resonance and noise shielding.

The main result is that these effects depend strongly on the initial preparation of the compound system and the placement of classical noise. The importance of this result could be appreciated in the cases where the behavior of a studied subsystem depends on the preparation of a bigger system whose parts have been traced out, but they are in interaction with the subsystem of interest. Our results are summarized as follows:

\begin {itemize}\item{ 2 qubits
\begin{enumerate}
\item{Stochastic anti-resonance is observed if the initial density matrix
contains population elements $\rho_{22}$ or $\rho_{33}$. This becomes evident in Bell states $|\Psi\rangle$, (\fref{2q1} (a) (upper curve)), while the entanglement of Bell states $|\Phi\rangle$ decays monotonically over increasing noise strength (\fref{2q1} (b))}.
\item{For vanishing dissipation rate ($\gamma\to 0$), the Hilbert subspace which exhibits stochastic anti-resonance tends to become a decoherence free subspace \cite{Ghikas}.}
\item{There are no noise shields in the 2-qubit case. Noise shields require the exten\-sion of the system to more qubits, in order to apply noise on the local fermionic environment.}
\item{Temperature degrades stochastic anti-resonance very quickly. An average exci\-ta\-tion number $\langle n\rangle\simeq0.5$ is enough to eliminate the appearance of the anti-resonance  (\fref{2q1} (a) (lower curve)).}
\end{enumerate}}

\item{ 3 qubits

\begin{enumerate}
\item{Multiple resonances (stochastic resonance and stochastic anti-resonance) are observed in two product states when noise affects the traced out qubit (\fref{3q} (a)). Moreover, there are two product  states that exhibit N.S. behavior , and two that exhibit S.AR. (\ref{a3q1}).}
\item{Product states with parallel spins do not present interesting behavior. Noise decreases monotonically their entanglement evolution. This holds for any ar\-ran\-gement of noise perturbation (\ref{a3q1}).}
\item{By altering the initial 3-qubit preparation of a given 2-qubit state, we observe different behaviour in the 2-qubit disentanglement time. This becomes evident with the reduced 2-qubit state of a W state. A small change of initial 3-qubit pre\-paration results in N.S., while  W state preparation results in S.AR. (\fref{3q2} (a,b)).}
\item{Bell state $|\Phi\rangle$ preparations exhibit N.S. when noise affects the traced out qubit, for all of the initial 3-qubit preparations (\fref{3q} (b) and \ref{a3q2}).}
\item{Most of the Bell state $|\Psi\rangle$ preparations exhibit N.S. when noise affects the traced out qubit, but there is an initial system preparation which results in S.R. (\fref{3q}(c) and \ref{a3q3}).}

\end{enumerate}}

\item{ 4 qubits

\begin{enumerate}
\item{Product states with parallel spins exhibit the same behavior as in the 3-qubit case (\ref{a4q1}).}
\item{Most of the product states  exhibit noise shield  when noise is environmental (\fref{4q1} (a) and \ref{a4q1}).}
\item{There are four product states which exhibit different kinds of effects, depending on the placement of noise and one that exhibits S.AR. behavior when noise is environmental (\ref{a4q1}).}
\item{The noise shields tend to become weaker if we increase the anisotropy of the system (\fref{4q1} (b)).}
\item{Most of the Bell state $|\Phi^+\rangle,|\Psi^+\rangle$ preparations  continue to exhibit N.S. when noise is environmental, except from two $|\Phi^+\rangle$ preparations that result in stochastic anti-resonance, two preparations of  $|\Psi^+\rangle$ which result in S.A.R., one $|\Psi^+\rangle$ preparation that exhibits multiple resonances and one $|\Psi^+\rangle$ preparation that exhibits S.R. (the last one when noise is internal). (\fref{4q12}(a) and \ref{a4q2}, \ref{a4q3}).}
\item{The noise shields become  stronger by applying noise in both of the traced out qubits rather than one of them (\fref{4q12} (b,c)). Temperature degrades their magnitude, but does not change their shape. There is increase of to\-lerance against temperature compared with the 2-qubit case, because of the environmental qubits. We can observe a clear N.S. behavior with excitation values up to $\langle n\rangle\simeq6$  (\fref{4q3}(a)).}
\item{Even though it is difficult to determine the aforementioned initial preparations which exhibit S.AR, we can control this behaviour by increasing temperature. Above a critical value of the latter, S.AR. disappears and we observe N.S.. This attributes an interesting positive role to the temperature, for the control of entanglement evolution (\fref{4q3}(b)).}
\end{enumerate}}
\end{itemize}

A complete list of the effects is presented in the tables of Appendix A and B. In the final paragraph we classify our results in terms of the observed effects and make comments for the interplay between the geometry of the initial system preparation and the placement of noise, which seems to be crucial for their appearance.
\section{XY Heisenberg model}
The general form of a N-spin XY chain (for spin $1/2$ particles) with nearest-neighbor interaction is
\begin{eqnarray}
H_{XY}=\sum_{n=1}^N\Big(J_xS_n^xS_{n+1}^x+J_yS_n^yS_{n+1}^y\Big),
\end{eqnarray}
with $S_n^i=\frac{1}{2}\sigma^i_n (i=x,y,z)$ the spin-$1/2$ operators, $\sigma_n^i$ the corresponding Pauli's ope\-rators, $\hbar=1$ and $S_{N+1}=S_1$ (periodic boundary condition). The chain is  ferromagnetic if $J_i<0$, and anti-ferromagnetic if $J_i>0$. Here we study this chain in the presence of a constant external magnetic field $\omega_0$ along the z-axis. Consequently, the unperturbed system Hamiltonian is:
\begin{eqnarray}
H_{0}=H_{XY}+\sum_{n=1}^N\omega_{0} S^z_n
\label{Hamiltonian},
\end{eqnarray}
for $N=3,4$.

The XY model has been studied for many decades because of its interesting and unusual features \cite{Wang}. It is an example of integrable system. While the iso\-tro\-pic Heisenberg chain is solved with the use of Bethe Ansatz, the XY model is solved by means of Jordan-Wigner transformation \cite{Jordan} which was introduced in $1928$ and applied to it by Lieb \etal \cite{Lieb} in $1961$. For more information see \cite{Parkinson}.  Furthermore, due to its mathematical simplicity, it is suitable for our purpose and  provides a basic unit for many quantum information implementations, such as N.M.R. quantum computation and quantum teleportation \cite{Rao,Yeo}.

\section{Markovian Master Equation}
We employ a Markovian form of the Master equation, where the classical gaussian white noise $\xi(t)$ affects the magnetic field and is appropriately incorporated in the double commutator \cite{Luczka}. For a thermal bath with $T\geq0$ the Master equation is
\begin{eqnarray}
\frac{d\rho_s}{dt}=&\nonumber-i[H_0,\rho_s]+\gamma(\langle n\rangle+1)\sum_{n=1}^N\Big(D[S^-_n]\rho_s\Big)\\&+\gamma \langle n\rangle\sum_{n=1}^N\Big(D[S^+_n]\rho_s\Big)-M_z[V_z,[V_z,\rho_s]],\label{master}
\end{eqnarray}
where $V_z$ is the spin operator affected by noise. It can be any of $S_n^z$ or sum of these. $M_z$ is the corresponding noise parameter. For example $M_4$ means that $V_4=S_4^z$, while $M_{34}$ means that $V_{34}=S_3^z+S_4^z$. Furthermore  $D[S^-_n]\rho_s=S^-_n\rho_s(S^-_n)^{\dagger}-\{(S^-_n)^{\dagger}S^-_n,\rho_s\}/2$. The coefficient $\gamma$ is the rate of population relaxation and $\langle n\rangle$ denotes the average excitation quanta of the bath. It depends monotonically on the temperature and it is used to parameterize the latter. For $T=0$ we have $ \langle n\rangle=0$ and for
$T\rightarrow\infty$, $ \langle n\rangle\rightarrow\infty$.
The first term describes the unitary evolution of the system, the second and third terms the interaction between the system and the thermal bath, while the last one is the addition of an external classical gaussian white noise \cite{Luczka}.
We assumed that each qubit has the same interaction with the environment, something that implies constraints on the value of coupling constants \cite{Ghikas,Wang}.
The eigen\-values of the 3-qubit Hamiltonian are:
\begin{eqnarray}\frac{{J+\omega\pm\sqrt{(J-2\omega)^2+3\Delta^2}}}{2}\\
\frac{{J-\omega\pm\sqrt{(J+2\omega)^2+3\Delta^2}}}{2}\\
\frac{-J\pm{\omega}}{2},\,\,\textnormal{double eigenvalues},
\end{eqnarray}
while the eigenvalues of the 4-qubit Hamiltonian are:
\begin{eqnarray}\pm\sqrt{{J}^{2}+{\Delta}^{2}+2{\omega}^{2}+\sqrt {{\Delta}^{4}+\left( 4{
\omega}^{2}+2{J}^{2} \right) {\Delta}^{2}+ \left( {J}^{2}-2{\omega}^{2}\right) ^{2}}}\\
\pm\sqrt{{J}^{2}+{\Delta}^{2}+2{\omega}^{2}-\sqrt {{\Delta}^{4}+\left( 4{
\omega}^{2}+2{J}^{2} \right) {\Delta}^{2}+ \left( {J}^{2}-2{\omega}^{2}\right) ^{2}}}\\
-J\pm\sqrt{\omega^2+\Delta^2},\\
J\pm\sqrt{\omega^2+\Delta^2}\\
\pm\omega\,\,\textnormal{double eigenvalues}\\
0\,\,\textnormal{fourfold eigenvalue}
\end{eqnarray}
where  $J=\frac{J_x+J_y}{2}$ and $\Delta=\frac{J_x-J_y}{2}$.
Assuming common $\gamma$ for all of the qubits, we work with $\omega=4$, $J=0.2$, $\Delta=0.1$, so that the interaction between them has not altered energy level separations by more than $10\%$ from that of the non-interacting system. Moreover, we choose $\gamma=0.01$ for the weak coupling approximation ($\omega\gg\gamma$) to remain valid.

\begin{figure}[ht]
\centering
\subfigure[$3$ qubits with a $2$-qubit subsystem]{\scalebox{0.3}{\includegraphics[scale=1.2]{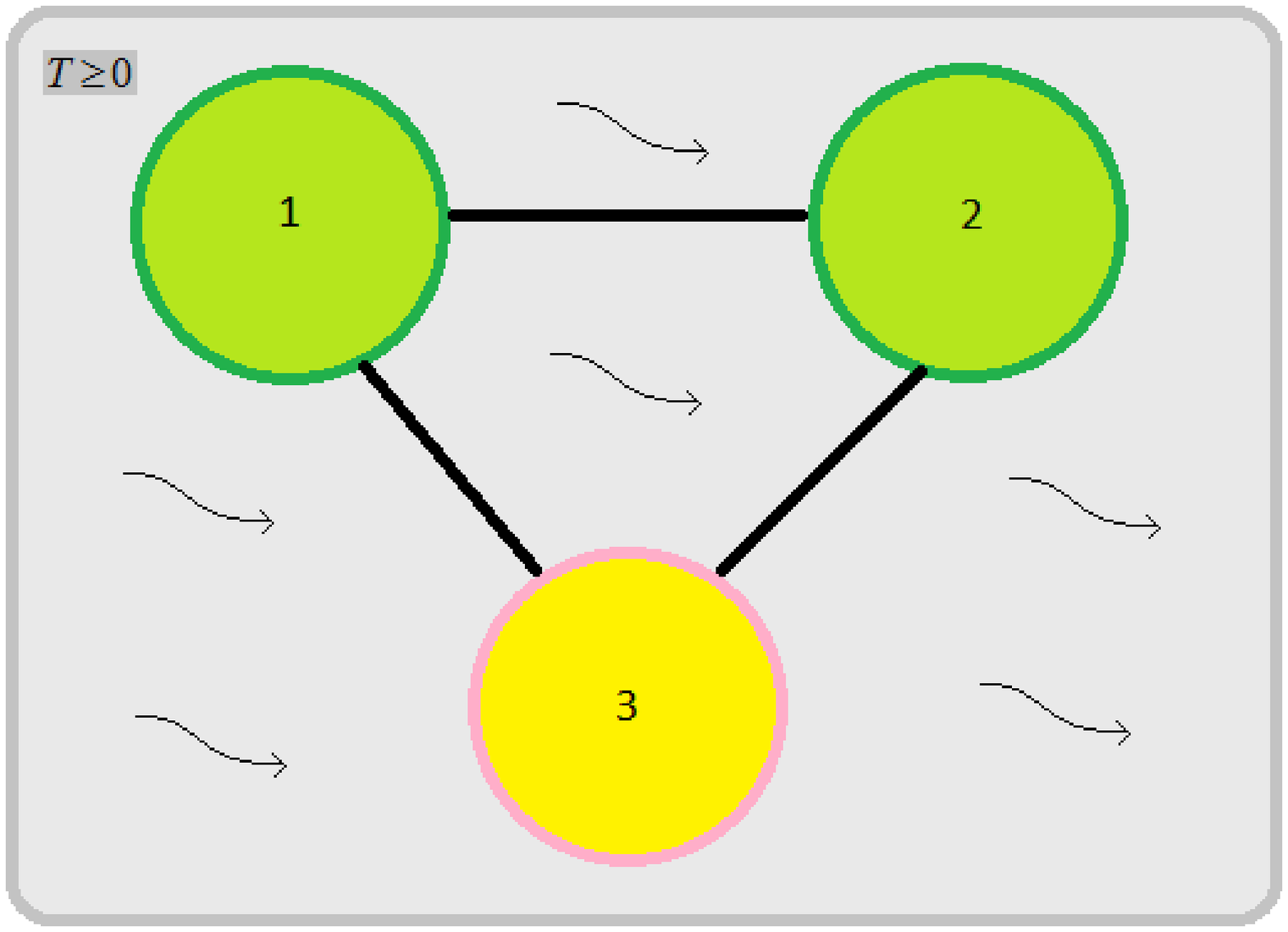}}}
\hspace{0.5in}
\subfigure[$4$ qubits with  two $2$-qubit subsystem]{\scalebox{0.3}{\includegraphics[scale=1.2]{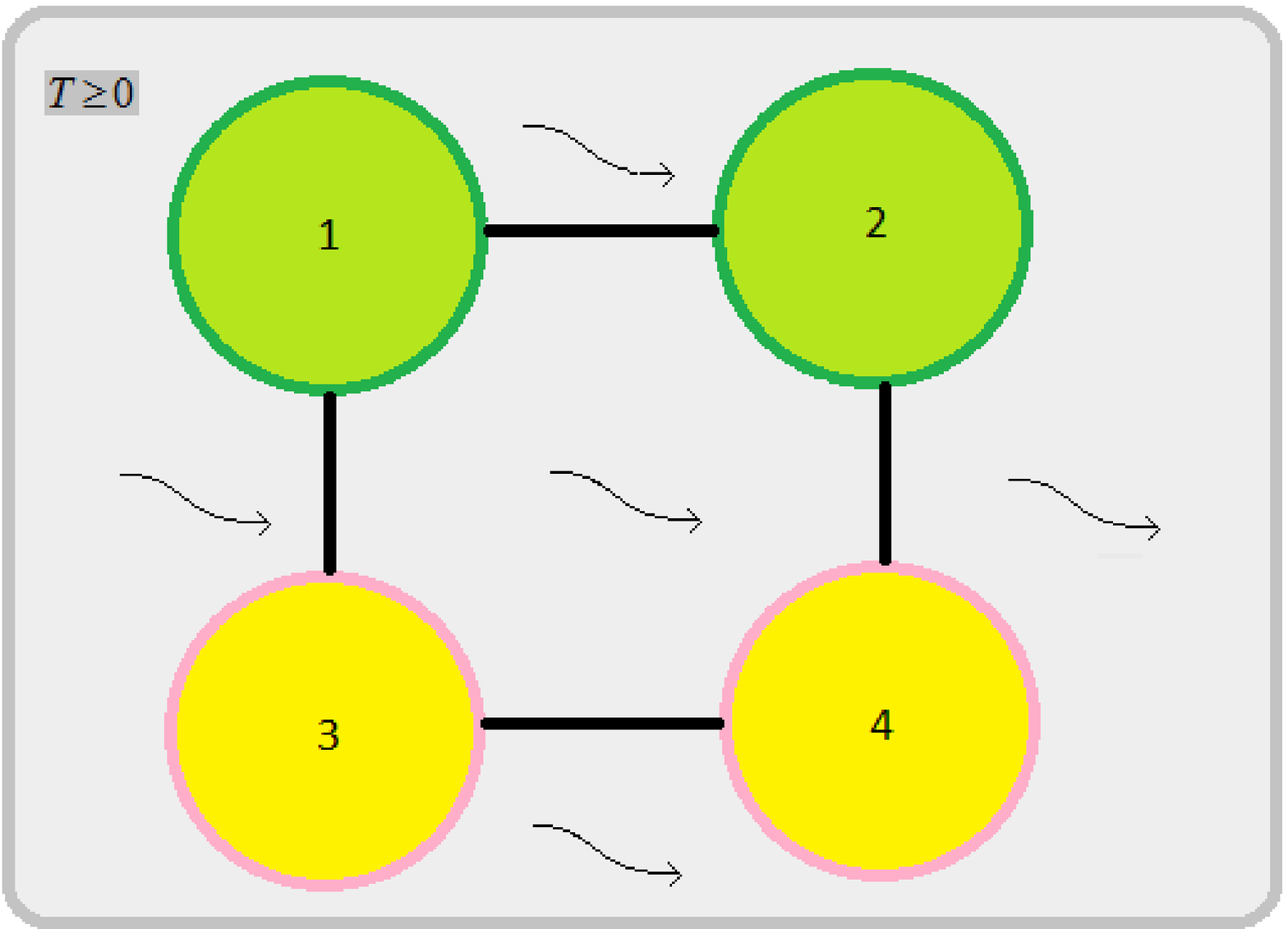}}}
\caption{The two different arrangements of the unperturbed system: $2$ qubit subsystem of a $3$-qubit chain (left) and a $4$-qubit chain (right) inside a common bosonic heat bath. The subsystem (green circles) interacts with the bosonic bath and the local fermionic environment (yellow circles). The main difference between them, is that in the case of $3$ qubits all of them interact directly, while in the case of $4$ qubits the interaction between $1-3$ and $2-4$ is indirect.}
\label{a1}
\end{figure}

\section{Entanglement Evolution}
Entanglement is a fundamental property of quantum systems and lies in the heart of quantum information and computation theory \cite{Nielsen,Horodecki}. One of the most difficult problems in contemporary information theory is to construct measures for the quantification of quantum entanglement. While the entanglement of a bipartite system is well understood, many measures have been presented for multipartite systems and most of them are related to pure states. This makes even more difficult the study of decohering systems, where initially pure states tend to become mixed, because of the constant interaction with their environment. That is why we choose to work with the bipartite entanglement, which is quantified by concurrence and gives us the advantage to describe both pure and mixed states \cite{Horodecki2,Kus,Wooters,Wooters2}.

For a system described by the density matrix $\rho$, the concurrence $C$ is

\begin{eqnarray}
C=\max(\sqrt{\lambda_1}-\sqrt{\lambda_2}-\sqrt{\lambda_3}-\sqrt{\lambda_4},0),
\end{eqnarray}
where $\lambda_1,\lambda_2,\lambda_3,\lambda_4$ are the eigenvalues of spin flipped density matrix R (with $\lambda_1$ the largest one), the definition of which is:
\begin{eqnarray}
R=\rho(\sigma_y\otimes \sigma_y)\rho^*(\sigma_y\otimes \sigma_y).
\end{eqnarray}
$C$ lies in the range $[0,1]$. $C=0$ corresponds to a product state, while $C=1$ to a maximally entangled state. All of the other states inside this range are  {\itshape partially entangled\/}.

The basis states for our system are formed by the tensor product of $S_n^z$ eigenstates $\{|e\rangle_n,|g\rangle_n\}$. In the 3-qubit case we have
\begin{eqnarray}
\{|eee\rangle,|eeg\rangle,|ege\rangle,|egg\rangle,|gee\rangle,|geg\rangle,|gge\rangle,|ggg\rangle\}
\end{eqnarray}
while in 4-qubit case
\begin{eqnarray}
\{|eeee\rangle,|eeeg\rangle,|eege\rangle,|eegg\rangle,|egee\rangle,|egeg\rangle,|egge\rangle,|eggg\rangle,\\|geee\rangle,
|geeg\rangle,|gege\rangle,|gegg\rangle,|ggee\rangle,|ggeg\rangle,|ggge\rangle,|gggg\rangle\}.
\end{eqnarray}

A simplifying feature of  our Hamiltonian  is that the dynamics of system splits into two independently evolving sets of equations, resulting  in smaller density matrices and consequently smaller differential systems. For three qubits case we get the density matrix \ref{d3} and for four qubits the density matrix \ref{d4} correspondingly. They both have the half the  number of elements of the original density matrix and result in the following reduced density matrix:
\begin{eqnarray}\rho^{r}=\left[ \begin{array}{cccccc}
\rho_{1,1}^{r} & & &\rho_{1,4}^{r} \\
 &\rho_{2,2}^{r}  &\rho_{2,3}^{r} & \\
 &\rho_{3,2}^{r}  &\rho_{3,3}^{r}  & \\
\rho_{4,1}^{r}  & & &\rho_{4,4}^{r}
\end{array}\right],\textrm{where blank entries are equal to $0$.}\end{eqnarray}
The mixed state defined by this submatrix is called  X state (non-zero elements along the diagonal and the anti-diagonal) and arises in many physical situations \cite{Rau,Rau2,James}. It has the advantage that includes Bell states $|\Psi^{\pm}\rangle, |\Phi^{\pm}\rangle$, product states and mixed states. The concurrence of this matrix is equal to:
\begin{eqnarray}
C=\max\{0,C_1,C_2\},
\end{eqnarray}
where
\begin{eqnarray}
&C_1=2(|\rho_{4,1}^{r}|-\sqrt{\rho_{3,3}^{r}\rho_{2,2}^{r}}),\\&
C_2=2(|\rho_{3,2}^{r}|-\sqrt{\rho_{4,4}^{r}\rho_{1,1}^{r}}),
\end{eqnarray}

It is evident that the reduced matrix for 2 qubits will inherit elements from the full system matrix which is either $8\times8$ (3-qubits case) or $16\times16$ (4-qubits case). This means that different initial preparations of the compound system can lead to the same reduced matrix form, something expected, but with great consequences for the entanglement evolution, as we will show in the next sections.

\section{2-qubit chain}
\begin{figure}[hb]
\centering
\subfigure[S.AR.]{\includegraphics[scale=0.25]{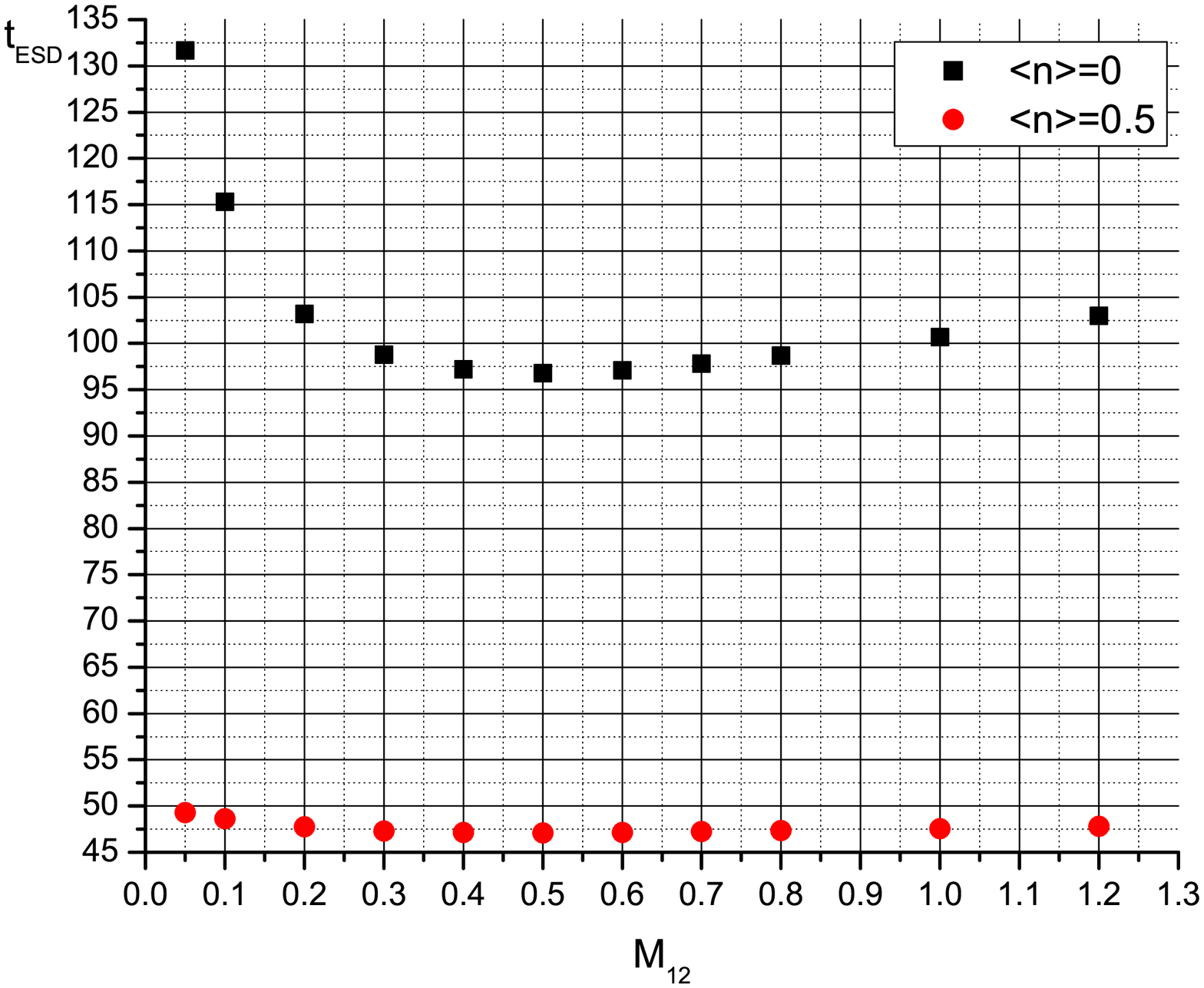}}
\subfigure[Monotonous Decay]{\includegraphics[scale=0.25]{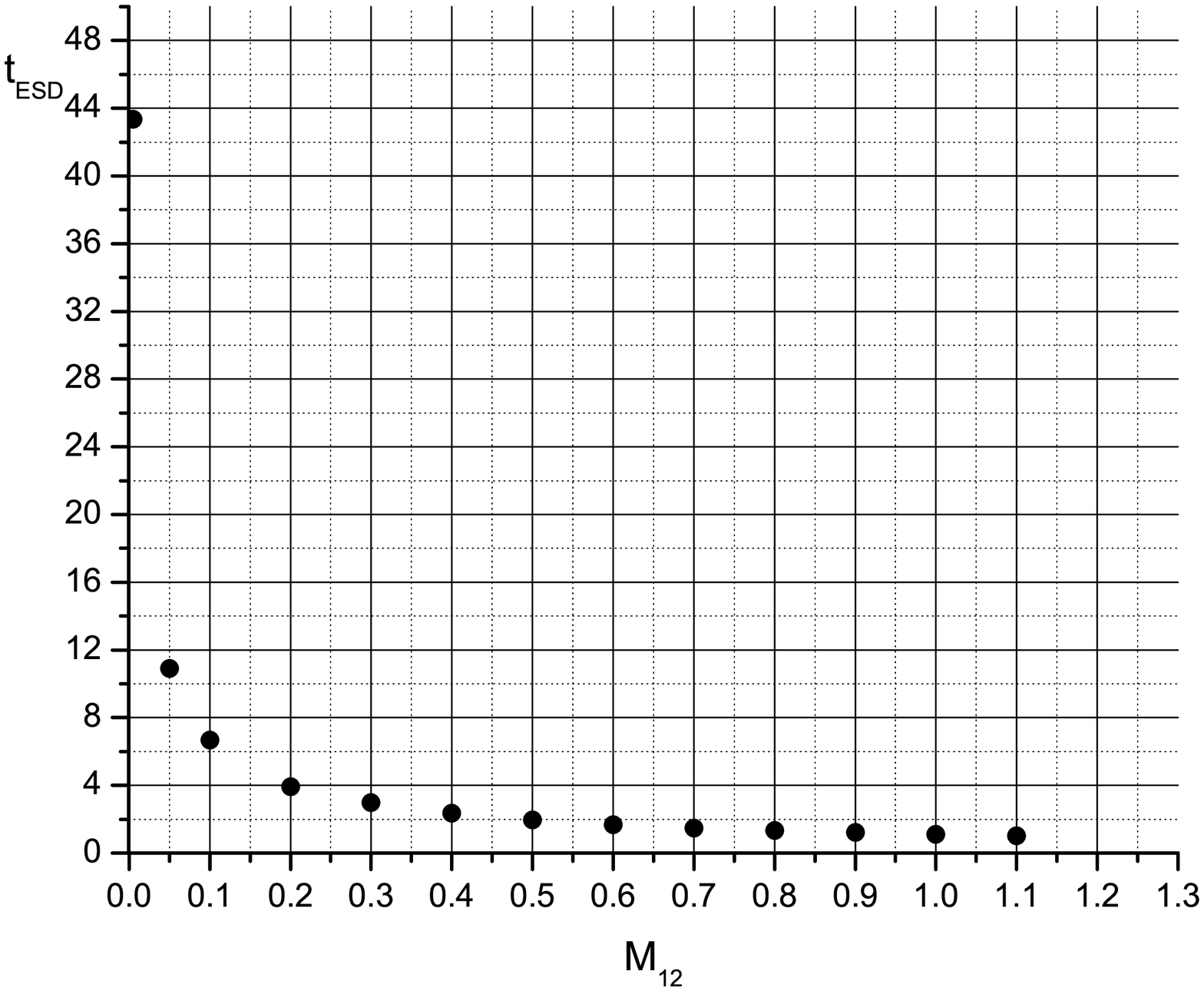}}
\caption{Entanglement sudden death time (E.S.D.) for $2$ interacting qubits, in the presence of a bosonic environment and external classical white noise. (a) Bell state $|\Psi^+\rangle$. The upper curve corresponds to zero temperature. There is a clear anti-resonance behaviour. The lower curve shows that a small increase of temperature erases the anti-resonance. (b) Bell state $|\Phi^+\rangle$ for $T=0$. There is no anti-resonance in this case. This shows that the geometry of the initial state is important for the appearance of this effect. Noise is collective.
($\omega = 1, \gamma = 0.01, J = \Delta = 0.1$ )}
\label{2q1}
\end{figure}

For the sake of completeness we provide two new diagrams related to those of \cite{Ghikas}. In the 2-qubit chain there is no ambiguity for the initial state preparation, since we do not trace out any qubit (we want to study entanglement). The system decoheres in the presence of a bosonic bath and external classical white noise affecting the magnetic field. What we found is that if the initial preparation of qubits contains the diagonal matrix elements $\rho_{22}$ or $\rho_{33}$, then the entanglement evolves non-monotonically over noise. There is a value of noise which causes the shortest disentanglement time and has to be avoided. This is what we call stochastic anti-resonance (S.AR.). The effect becomes evident in the Bell states $|\Psi\rangle$ (\fref{2q1} (a) (upper curve)). Bell states $|\Phi\rangle$ decay monotonically, since their evolution begins outside the critical subspace (\fref{2q1} (b)). Regarding the bath temperature, it plays a negative role for the S.AR., something normally expected. A small increase of temperature is enough to eliminate the shape of S.AR. (\fref{2q1} (a) (lower curve)).

\section{3-qubit chain with $T=0$}
The reduced density matrix has the general form:
\begin{eqnarray}\label{3q1}
\rho^{r}=\left[ \begin{array}{cccccc}
\sum_{i=1}^{2}\rho_{i,i}&0&0&\sum_{i=1}^{2}\rho_{i,i+6}\\ 0&\sum_{i=3}^{4}\rho_{i,i}&\sum_{i=3}^{4}\rho_{i,i+2}&0\\0&\sum_{i=3}^{4}\rho_{i+2,i}& \sum_{i=5}^{6}\rho_{i,i}&0\\ \sum_{i=1}^{2}\rho_{i+6,i}&0&0&\sum_{i=7}^{8}\rho_{i,i}
\end{array}\right]
\end{eqnarray}
Obviously it inherits elements from the 3-qubit density matrix \ref{d3}. Different prepa\-rations of the latter result in the same state of the subsystem. When all of the elements of the compound density matrix that participate in the sums of the reduced density matrix are equal, we call the state ``balanced'' (see the first preparation of Bell states in \ref{a3q2}, \ref{a3q3}, \ref{a4q2}, \ref{a4q3}).

Firstly we focus on product states of 3-qubits which result, obviously, in 2-qubit product states. In order to study the entanglement evolution of product states,  we calculated the area included between the concurrence graph and the time axis until concurrence reaches zero for the first time after its production, for different values of noise strength. We are interested here in the first cycle of creation-decay of entanglement, which has the largest amount of the latter, and not for any rebirths. This measure is sensitive in all kinds of behavior the entanglement production-decay can exhibit (smooth or oscillatory) and carries information for both of the maximum of entanglement production and the disentanglement time.
The results are presented in the \tref{a3q1}. There are two non-interesting cases where all qubits are aligned ($|eee\rangle$ and $|ggg\rangle$) and noise degrades entanglement monotonically regardless of its placement, two states which present S.AR. behaviour, two states with multiple resonances (S.R. and S.AR.)  and two with N.S. behavior. The results show clearly that all of the above effects occur when the noise is applied on the environmental qubit ($M_z=M_3$).

\begin{figure}[hb]
\centering
\subfigure[S.R. and S.AR.]{\scalebox{0.25}{\includegraphics{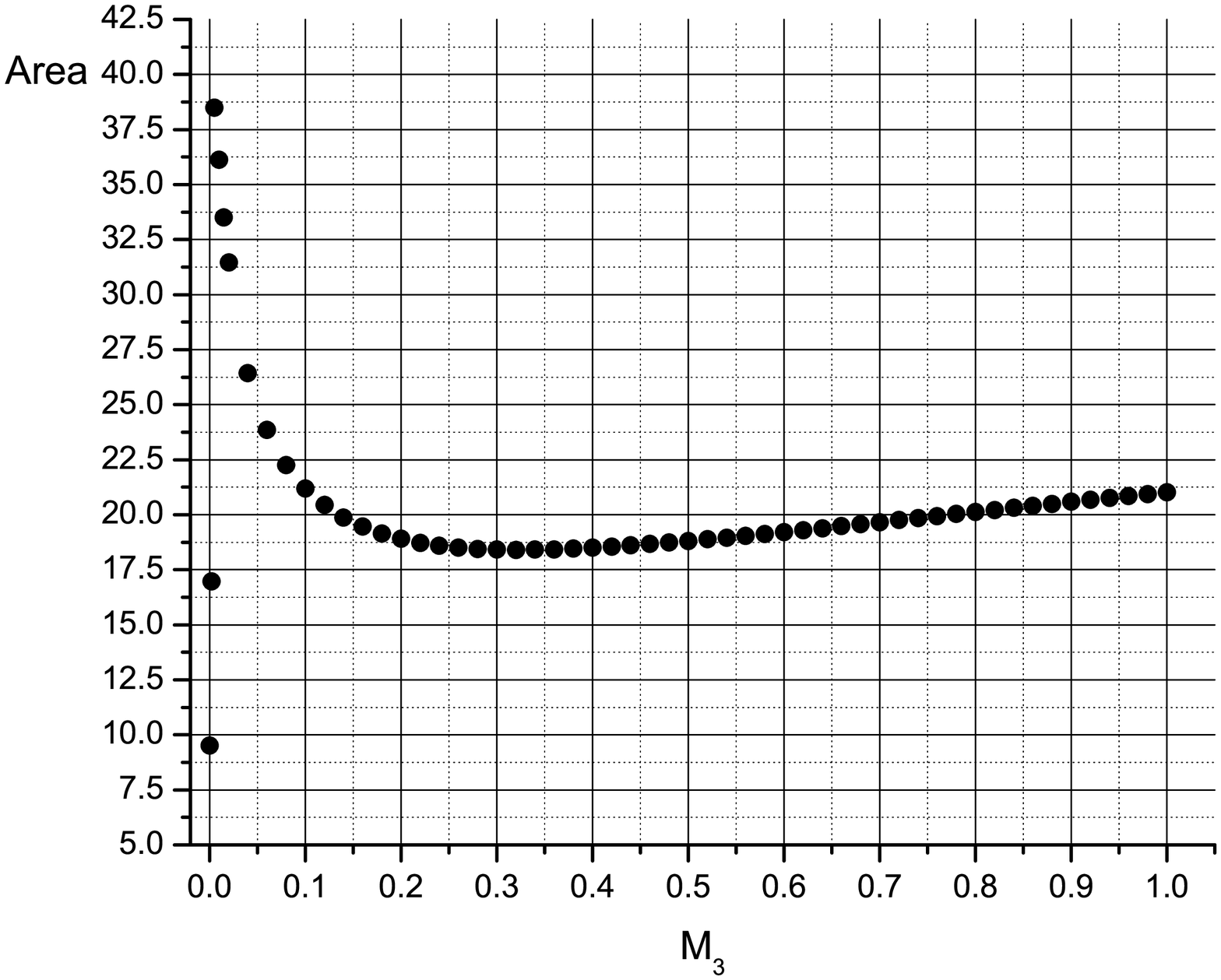}}}
\subfigure[N.S.]{\scalebox{0.25}{\includegraphics{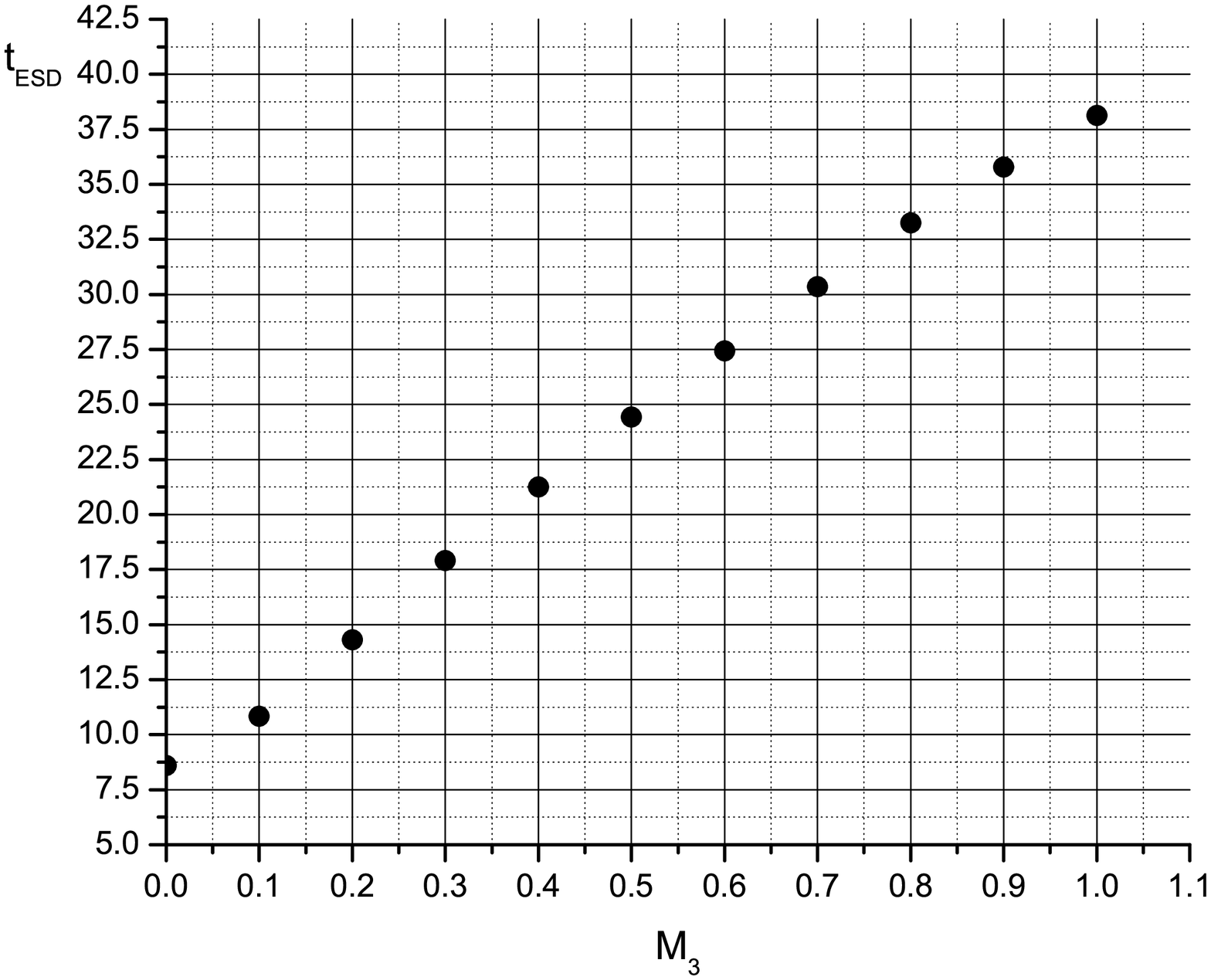}}}
\subfigure[N.S. and S.R. comparison]{\scalebox{0.25}{\includegraphics{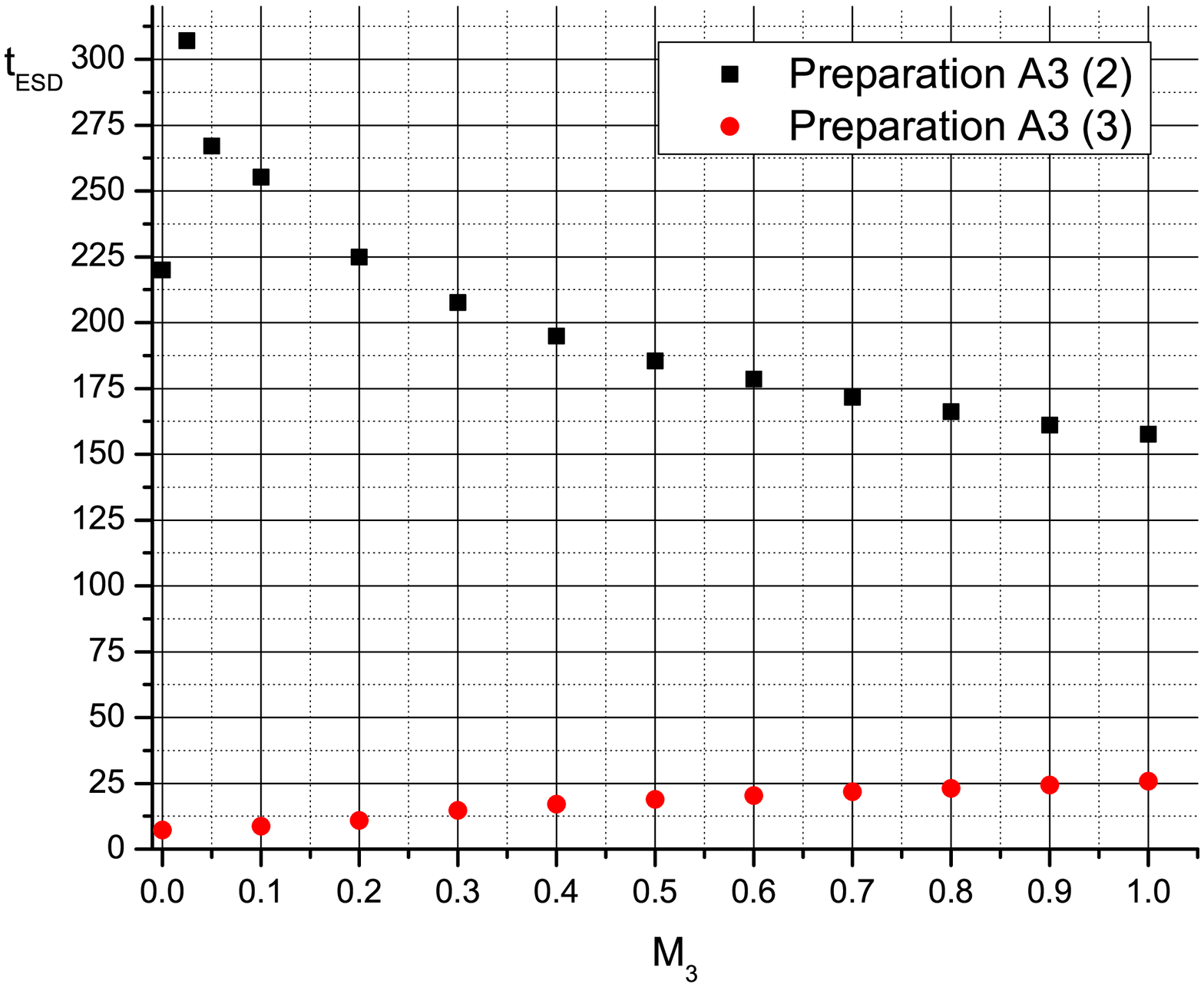}}}
\caption{3 interacting qubits. E.S.D. time for several initial states, as a function of  noise parameter. (a) We observe the multiple resonances (S.R and S.AR.) for the initial product state $|geg\rangle\to|ge\rangle$, where the S.R is very sharp and occurs with a very small noise strength, while S.AR. is much smoother and occurs with a larger noise strength. (b) N.S. behavior of the balanced Bell state $|\Phi^+\rangle$. (c) Two different preparations of Bell state $|\Psi^+\rangle$ that exhibit S.R. and N.S. correspondingly. Noise is external (affects the traced out qubit) in all of the above cases. ($\langle n\rangle=0$, $\omega_0=4$, $J=0.2$, $\Delta=0.1$, $\gamma=0.01$)}\label{3q}
\end{figure}
Then we move on to the 2-qubit Bell states ${|\Phi^+\rangle}, {|\Psi^+\rangle}$ and try several initial 3-qubit preparations that result in them.
Almost all of them exhibit noise shield behaviour when noise lies outside our subsystem except from one. Indeed, there is a preparation of $|\Psi^+\rangle$ which exhibits stochastic resonance when noise is environmental (\ref{a3q2}, \ref{a3q3}, \fref{3q} (b,c)).

\begin{figure}[hb]
\centering
\subfigure[S.AR.]{\scalebox{0.25}{\includegraphics{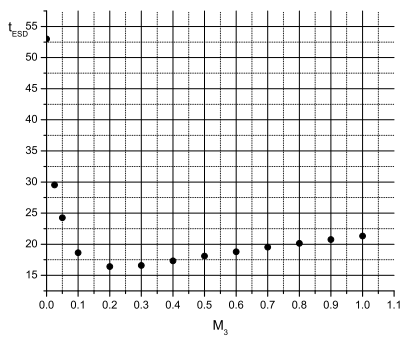}}}
\subfigure[N.S.]{\scalebox{0.25}{\includegraphics{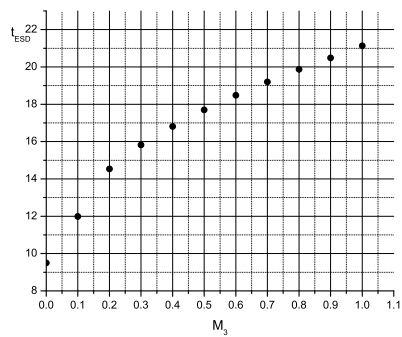}}}
\caption{3 interacting qubits. (a) E.S.D. time for the state resulting after tracing out the third qubit of the W state $\frac{|eeg\rangle+|ege\rangle+|gee\rangle}{\sqrt{3}}$, as a function of external noise parameter $M_{3}$. We observe S.AR behavior. (b) By altering a little bit the 3-qubit preparation we get a completely different behavior, a N.S.. For small values of noise, the disentanglement times are quite smaller than those of the previous case. Noise is external (affects the traced out qubit) in all of the above cases.   ($\langle n\rangle=0$, $\omega_0=4$, $J=0.2$, $\Delta=0.1$, $\gamma=0.01$)}\label{3q2}
\end{figure}

Finally we present our results for the bipartite subsystem of W state
\begin{eqnarray}
\frac{|eeg\rangle+|ege\rangle+|gee\rangle}{\sqrt{3}},
\end{eqnarray}
 which has been proposed as the maximum 3-qubit entangled state and has non-zero entanglement across any bipartition.  The  W state has matrix representation
\begin{eqnarray}\rho_{W}=\left[ \begin{array}{cccccccc}
 0&0 &0 &0 &0&0&0&0\\
 0&\frac{1}{3} &\frac{1}{3}&0&\frac{1}{3}&0&0&0 \\
 0&\frac{1}{3} &\frac{1}{3} &0&\frac{1}{3}&0&0&0 \\
 0&0  &0  &0&0&0&0&0 \\
 0&\frac{1}{3} &\frac{1}{3} &0&\frac{1}{3}&0&0&0 \\
 0&0 &0 &0&0&0&0&0 \\
  0&0 &0 &0&0&0&0& 0\\
 0&0 &0 &0&0&0&0&0
\end{array}\right]\end{eqnarray}
and its  reduced $2$-qubit density matrix  is
\begin{eqnarray}\rho^{r}=\left[ \begin{array}{cccccccc}
 \frac{1}{3}&0 &0 &0 \\
 0&\frac{1}{3} &\frac{1}{3}&0 \\
 0&\frac{1}{3} &\frac{1}{3} &0\\
 0&0  &0  &0
\end{array}\right],\end{eqnarray}
with initial entanglement $C_{initial}\simeq0.667$. Again, we observe S.AR. only when the noisy qubit is  environmental. The most interesting result, though, is that if we prepare the same $2$-qubit state, but with different elements of the $3$-qubit density matrix, we observe N.S. behavior (\fref{3q2} (a,b)). This can be done easily by setting  $\rho_{2,3}=\rho_{3,2}=0$ in the W state preparation. The reduced density matrix (\ref{3q1}) does not contain these elements, so the $2$-qubit state remains the same. The ambiguity in the preparation of the subsystem has important consequences on its evolution.

\section{4-qubit chain with $T=0$}
The reduced density matrix in this case is:
$$\rho^{r}=\left[ \begin{array}{cccccc}
\sum_{i=1}^4\rho_{i,i}&0&0&\sum_{i=1}^{4}\rho_{i,i+12}\\ 0&\sum_{i=5}^{8}\rho_{i,i}&\sum_{i=5}^{8}\rho_{i,i+4}&0\\0&\sum_{i=5}^{8}\rho_{i+4,i}& \sum_{i=9}^{12}\rho_{i,i}&0\\ \sum_{i=1}^{4}\rho_{i+12,i}&0&0&\sum_{i=13}^{16}\rho_{i,i}
\end{array}\right]$$

The main difference from the 3-qubit case is that the qubits do not interact directly with each other.Qubit  $1$ does not interact directly with $3$ (same for $2$ and $4$). We continue to study the entanglement between $1$ and $2$, because we are interested in the behavior of two directly interacting qubits.

\begin{figure}[hb]
\centering
\subfigure[N.S. for a product state]{\scalebox{0.25}{\includegraphics{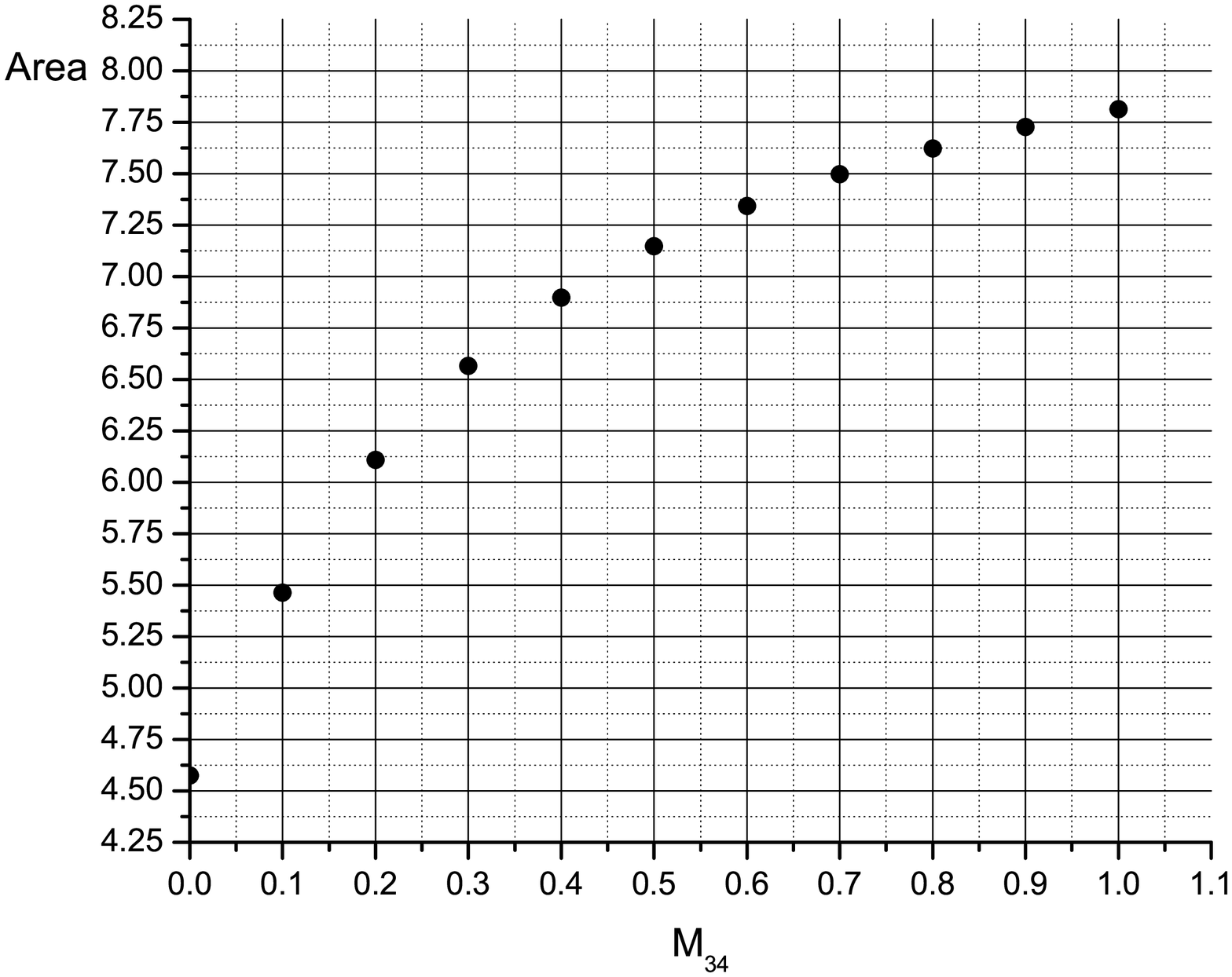}}}
\subfigure[N.S. for different $\Delta$]{\scalebox{0.25}{\includegraphics{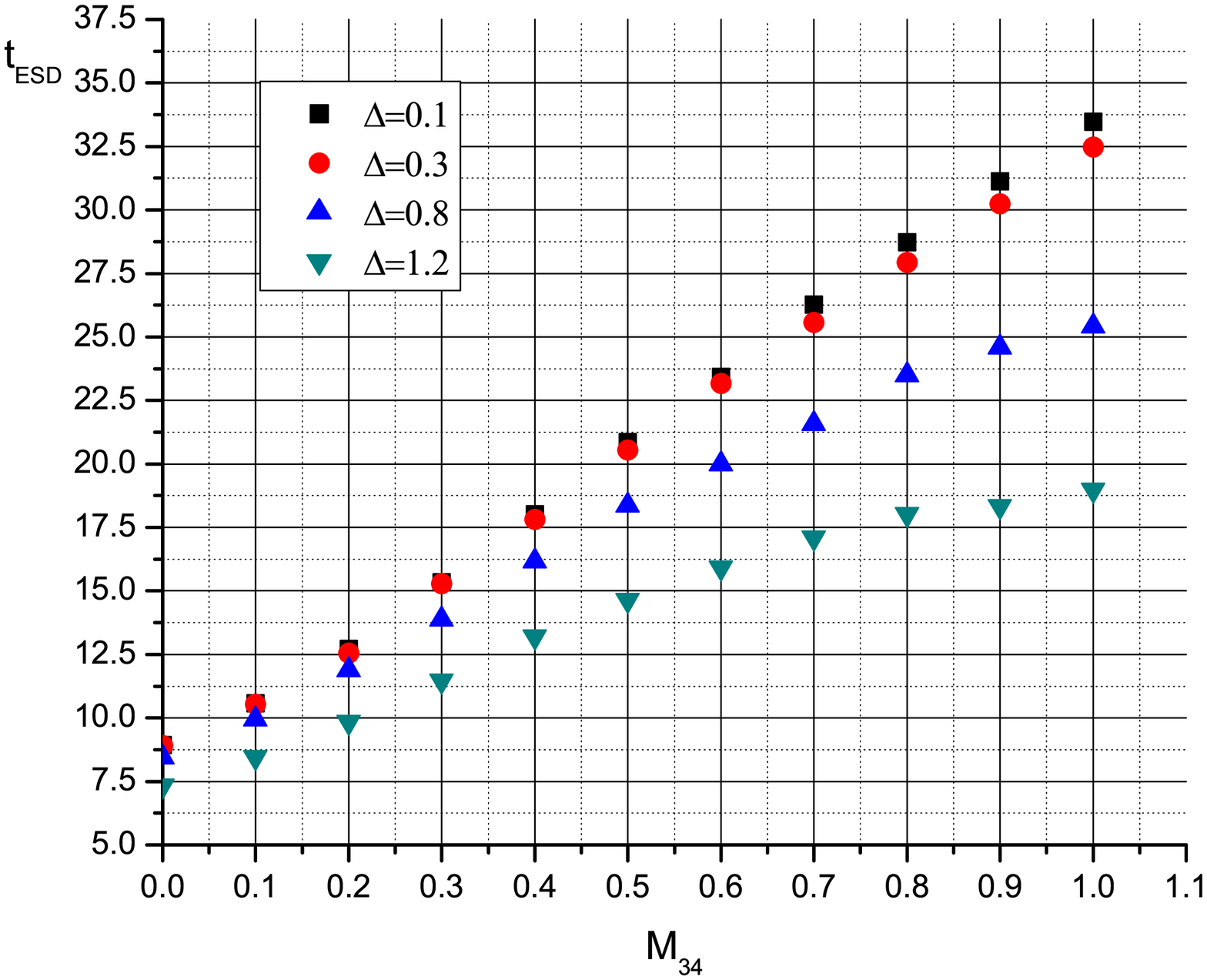}}}
\caption{4 interacting qubits (a) N.S. for the product state $|eggg\rangle\to|eg\rangle$. (b) N.S. of the balanced $|\Phi^+\rangle$ against anisotropy constant $\Delta$. The bigger the anisotropy, the weaker the noise shield. ($\langle n\rangle=0$, $\omega_0=4$, $J=0.2$, $\Delta=0.1$, $\gamma=0.01$)}\label{4q1}
\end{figure}

As we can see in \tref{a4q1}, there are many product states where the combination of entanglement production and  disentanglement time increases  monotonically over noise, when  the latter is environmental. This is noise shield behavior (\fref{4q1} (a)). We also observe three cases where multiple resonances become evident. Two of them ($|ggeg\rangle, |ggge\rangle$) ``do not obey the rule'' and exhibit resonances only when the noise penetrates at least one qubit of the subsystem under study.

The behavior of Bell states $|\Phi^+\rangle$ and  $|\Psi^+\rangle$ is not the same any more. While in the 3-qubit case Bell states $|\Phi^+\rangle$ exhibit N.S. for all of the initial preparations we have studied, here there are some initial preparations that cause the system to exhibit stochastic anti-resonance, similarly to the W state case of 3-qubits (\ref{a4q2}). Furthermore, we found that the N.S. becomes weaker when the anisotropy parameter $\Delta$ increases (\fref{4q1} (b)).

\begin{figure}[hb]
\centering
\subfigure[E.S.D. for different preparations of $|\Psi\rangle$]{\scalebox{0.25}{\includegraphics{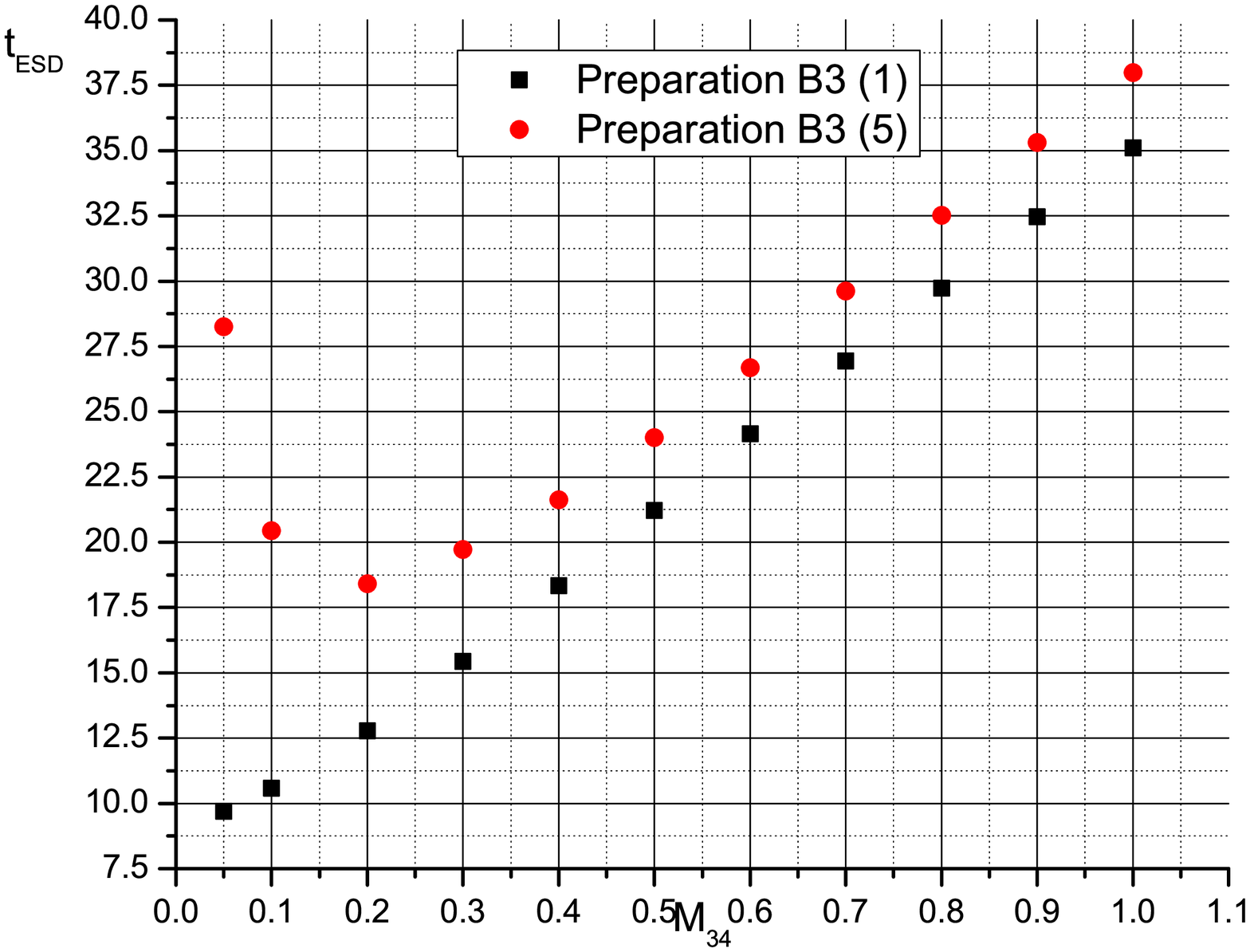}}}
\subfigure[Noise affecting the subsystem]{\scalebox{0.25}{\includegraphics{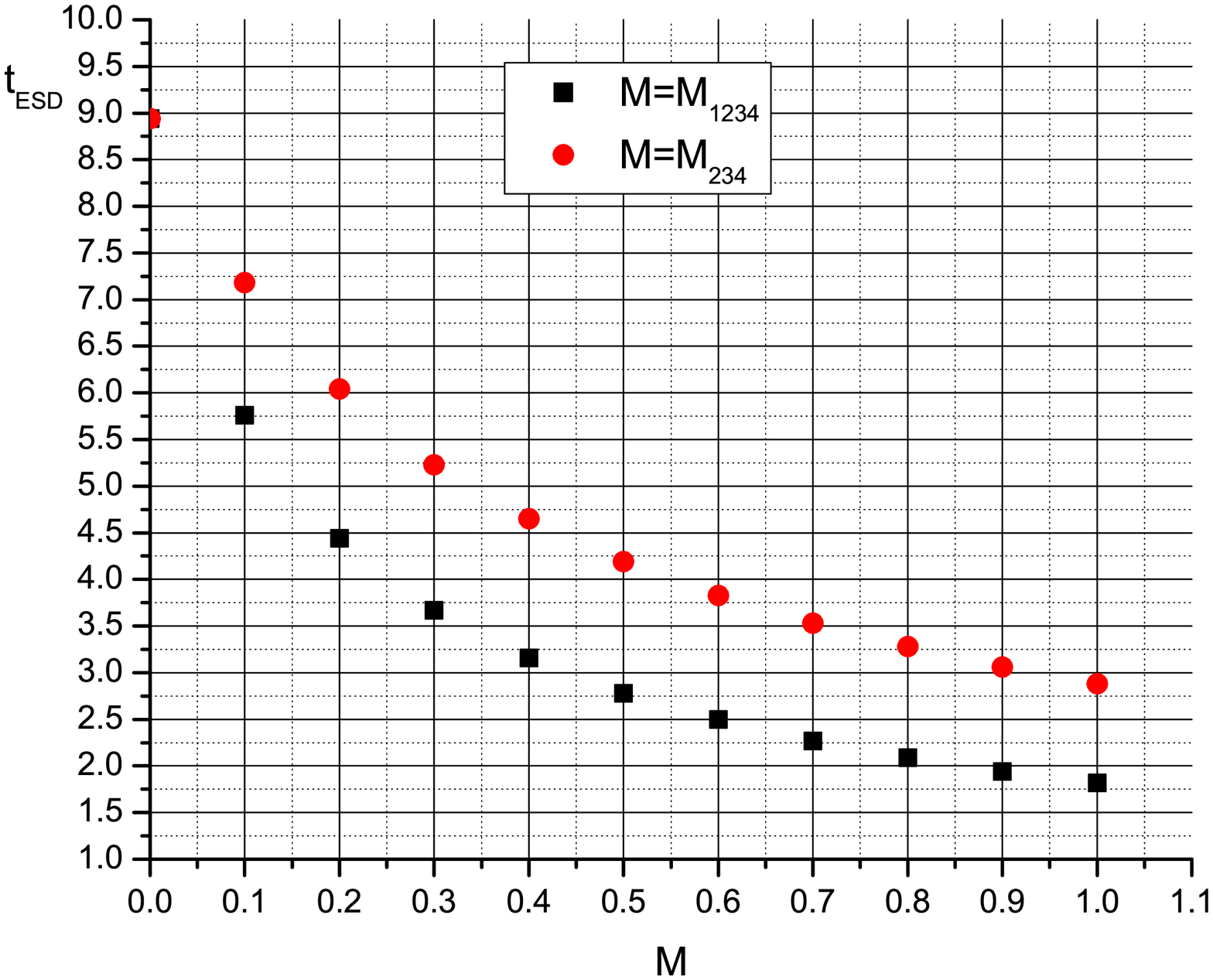}}}
\subfigure[N.S. comparison for different placements of classi\-cal noise]{\scalebox{0.22}{\includegraphics{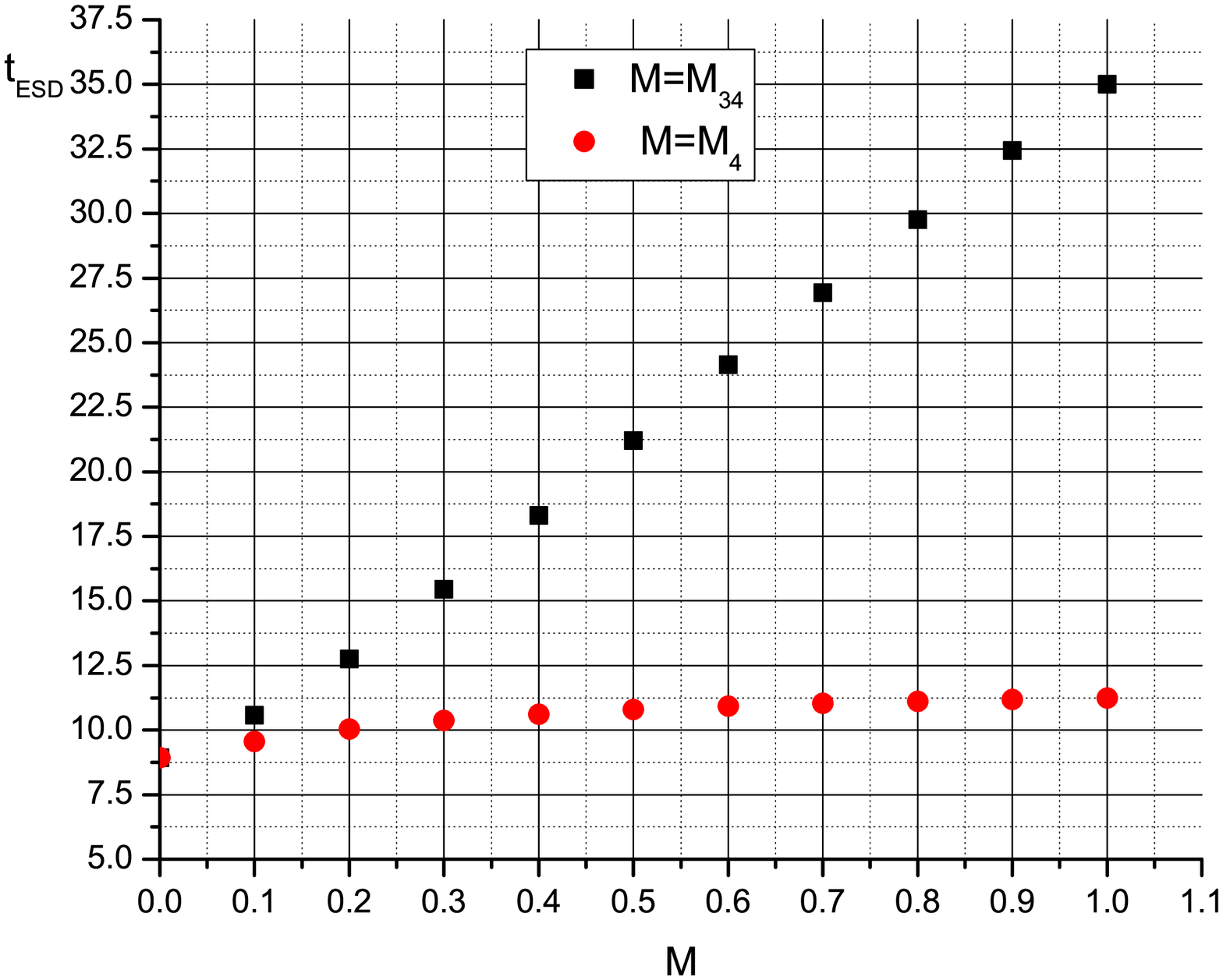}}}
\caption{4 interacting qubits. (a) E.S.D. time for two preparations of $|\Psi^+\rangle$, when noise affects the two traced out qubits. The black squares correspond to the first preparation (balanced state) in \tref{a4q3} and the red dots to the fifth one. It is clear that the initial state subspace determines the entanglement evolution: The first preparation exhibits N.S, while the second one exhibits S.AR. and is characterized by much longer disentanglement times for small values of noise. (b) E.S.D. time for the balanced Bell state $|\Psi^+\rangle$, as a function of external noise parameter. The system response is better when noise affects only one ``internal'' qubit and not both of them, something expected. (c) We observe a substantial change of behavior when placing noise out of the subsystem. The response is monotonically positive (i.e. N.S.). We find a better behavior (almost linear), with much longer disentanglement times, when both of the environmental qubits are noisy. The same hold for all  of the Bell state $|\Phi^+\rangle$ preparations. ($\langle n\rangle=0$, $\omega_0=4$, $J=0.2$, $\Delta=0.1$, $\gamma=0.01$)}\label{4q12}
\end{figure}

Bell state $|\Psi\rangle^+$ presents a variety of effects depending on its initial preparation and the placement of noise. We found N.S., S.AR. and S.R. (\ref{a4q3} and \fref{4q12} (a)).
In \fref{4q12} (b) and \fref{4q12} (c) we present the results for the balanced Bell state $|\Psi^+\rangle$. It becomes evident that the application of noise should take place out of the subsystem, in order to create a noise shield. Moreover, the later is amplified when both environmental qubits are noisy, rather than one, something intuitively expected.

\section{Non-zero temperature and special initial Hilbert subspaces}
As we presented above,  there are some initial preparations both in 3 and 4-qubit chains that cause the subsystem to exhibit S.AR or S.R., while other result in N.S. behavior, which is obviously the best result we can get by adding noise to a quantum system.

The identification of these ``special'' regions of initial Hilbert space is a very difficult task, because of the great number of real parameters needed to describe the  density matrix of a 3-qubit or 4-qubit system. The XY interaction Hamiltonian and our Master equation may simplify significantly the system of differential equations we need to solve, but even in this case, the number of real independent parameters for $3$ qubits is $31$ while for $4$ qubits is $127$.

Moreover, the choice of working on the bipartite entanglement does not simplify the problem: The evolution of the reduced 2-qubit density matrix has direct dependance on the non participating elements of the full density matrix, something expected, because one solves firstly the full differential system and then traces out the desired number of qubits.
Needless to say that if we wish to study larger systems, then the problem acquires extremely large complexity and becomes practically unsolvable.

There are two ways to approach this problem:
\begin{enumerate}
\item{{\it The theoretical one}: We have to understand how the initial geometry of the full system provides the necessary background for the beneficial (or not) synchronization of two coexisting timescales (dissipation and decoherence). This could lead to the optimized exploitation of noise addition in order to get the best possible result, which is obviously a noise shield. We believe that the solution of this problem will come from a deeper analysis of the geometry of quantum states and the multipartite entanglement measures.}
\item{{\it The practical one}: Given the complexity of the problem, it would be helpful if we could control the extent of these areas by means of an external macroscopical variable.}
\end{enumerate}

\begin{figure}[hb]
\centering
\subfigure[E.S.D. time for a Bell State]{\scalebox{0.25}{\includegraphics{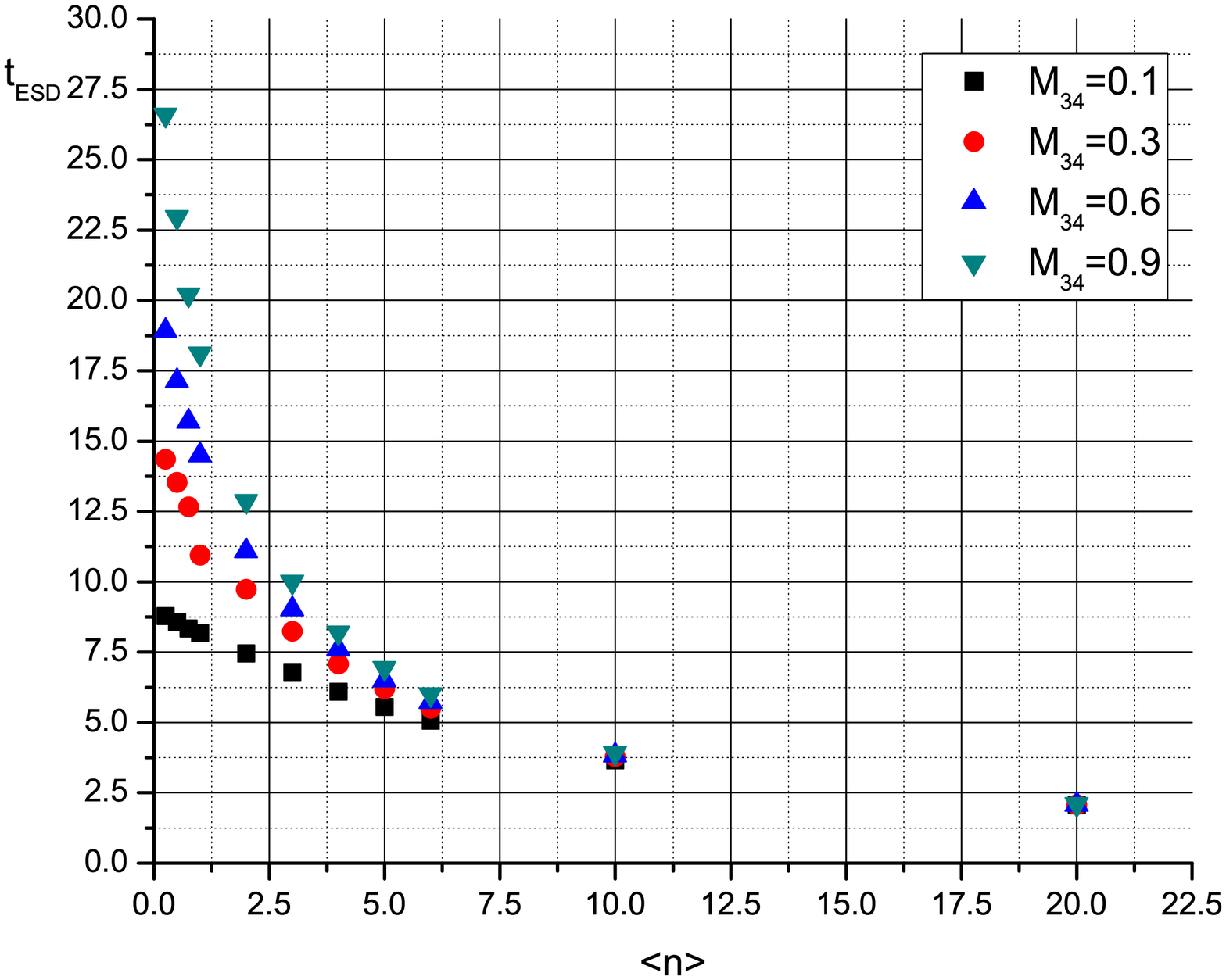}}}
\subfigure[S.AR. and N.S. for special preparations]{\scalebox{0.25}{\includegraphics{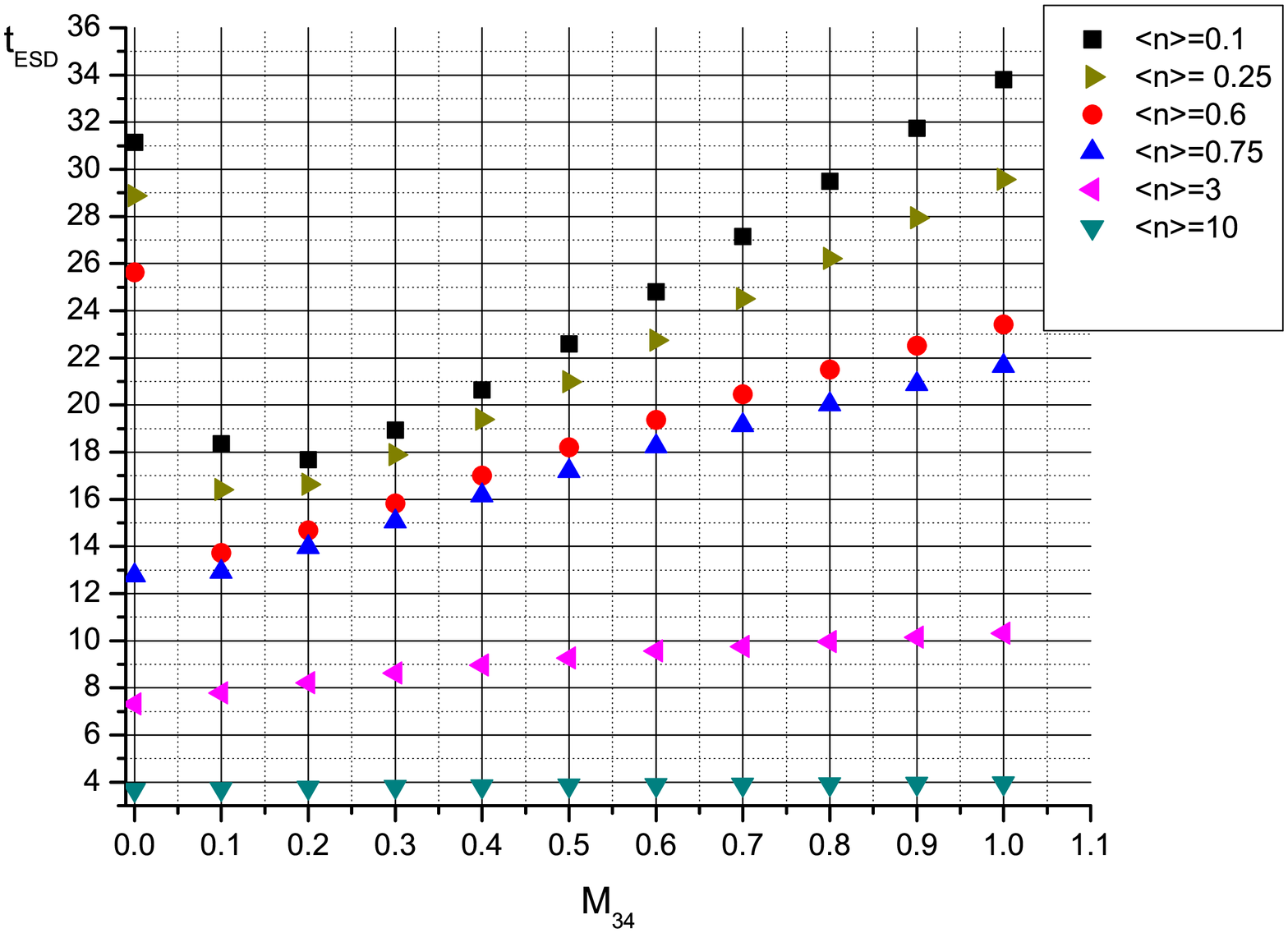}}}
\caption{4 interacting qubits (a) E.S.D time for the balanced Bell state $|\Psi^+\rangle$ as a function of temperature parameter $\langle n\rangle$, for different values of $M_{34}$. We observe the robustness of N.S. against temperature, although the absolute value of $t_{ESD}$ becomes smaller as $\bar{n}$ gets bigger (something expected). (b) Another preparation of $|\Psi^+\rangle$ (\ref{a4q3} (5)) which exhibits S.AR.. It is clear that above a critical $\langle n\rangle$ we get N.S. instead of S.AR., something that attributes a positive role to the bath temperature. ($\omega_0=4$, $J=0.2$, $\Delta=0.1$, $\gamma=0.01$)}\label{4q3}
\end{figure}

In the present stage of our understanding of the problem of which classes of states are related to noise induced effects, we think that finding an external control seems more straightforward. One option is to study the effect of temperature on the entanglement evolution, since temperature has the advantage of being an easily controllable macroscopical variable.
In our previous work \cite{Ghikas} we observed the negative role of a small increase of temperature in two qubits of XY model (\fref{2q1} (a)). By moving to a larger system ($3$ and $4$ qubits), we expected to see shielding effects, due to the environmental qubits.

This was indeed our first result: The larger the fermionic environment we have, the bigger the tolerance of the subsystem against noise. Although the disentanglement time decreases as $\langle n\rangle$ gets bigger, the results are the same for all of the states which exhibit monotonic behavior in $T=0$ case: Noise shield for $T=0$ remains shield for $T>0$ as we can see in \fref{4q3} (a). We note again that in the case of $4$ qubits the shield becomes ``stronger'' when both of the external qubits are noisy.

The most interesting result though, is the effect of temperature increase on those special initial preparations that result in S.AR. behavior. In the 4-qubit case there is a $\langle n\rangle_{critical}$ under which S.AR. becomes evident. Above this value, the subsystem acquires N.S. behavior as we can see in \fref{4q3} (b). Temperature eliminates the first-negative part of S.AR. and leaves  only the second-positive part (ie. N.S.). This is a very welcome result, because it gives us the opportunity to get the best possible noise effect, the noise shield, whether the subsystem begins from a ``S.AR. region'' or not and provides a way to avoid the identification of the latter: If we observe S.AR. then we must increase the temperature and get N.S.!

\section{Conclusions}

In our study of a system of two interacting qubits, under the influence of a bosonic environment and classical noise, we had observed that the appearance of effects of non-monotonic dependence  of the disentanglement time on the noise strength, was dependent on the initial preparations of the system \cite{Ghikas}. Then the natural question that arose was whether the initial preparations of the local environment of the 2-qubit subsystem does influence the dynamical evolution. As it has turned out, the local fermionic environment does indeed influence the way the subsystem reacts to an external classical noise.
We have studied several initial preparations of 3-qubit and 4-qubit chains and focused on the bipartite entanglement of the reduced subsystem. Even though they differ by only one qubit, they represent two different cases of a quantum system: In the $3$-qubit chain all qubits interact directly with each other as we see in \fref{a1}.  This is not the case in general  for Heisenberg type $n$-qubit chains with $n\geq4$. Our system is modelled by a Markovian Master equation with two Lindblad dissipators: one representing the common quantum bath and one representing the classical external field. While the first one is related to the amplitude damping channel, the second one implies a continuous indirect measurement of $V_z$. In \cite{Breuer} one can see that the last term of the Master equation points to the quantum Zeno effect \cite{Misra}, which has been proposed as a reliable method for the protection of entanglement against environmental noise \cite{Zaffino,Francica}.

What we have observed, is that the application of external noise in combination with the proper initial preparation of the compound system, has clear influence on  the entanglement evolution of the subsystem.
While many initial preparations of a compound system result in the same subsystem state, they may lead to  very different dynamical behaviours of the subsystem entanglement.
The $3$ interesting effects we observed, are:
\begin{itemize}
\item{{\it Noise shield}: Monotonous increase of disentanglement time against noise. This is the most likely effect, according to our summary tables, when noise is environmental (both in $3$ and $4$-qubit chains). The traced out qubit(s) play the role of a local fermionic environment which shields the subsystem against dissipation and decoherence. It is strongly related to quantum Zeno effect.}
\item{{\it Stochastic anti-resonance}: Non-monotonic behavior of the disentanglement time, firstly decreasing and then increasing. This means that a moderate value of noise has the worst possible effect on the system and needs to be avoided. This was initially observed in our 2-qubit study \cite{Ghikas} for various initial states. In our 3 and 4 qubit study, the effects depend not only on the preparation but also on the way the noise is applied.  We observed  S.AR. for $4$ initial preparations  in the 3-qubit chain case, with noise acting only on the traced out qubit. In the 4-qubit case, there were $9$ appearances of S.AR., $7$ with noise acting only on the environmental qubits and $2$ with noise inside the subsystem.}
\item{{\it Stochastic resonance}: Non-monotonic behavior of the disentanglement time, firstly increasing and then decreasing. There is a moderate value of noise that upgrades system reaction. We observed it $3$ times in 3-qubit chains, with noise acting on the traced out qubit and $5$ times in 4-qubit chains, $3$ with noise inside the subsystem and $2$ with noise out of the subsystem.}
\end{itemize}

We note that  product and entangled states have different time evolution. Product states begin at zero entanglement, they evolve to entangled states because of qubit interaction and finally loose their entanglement in the presence of environmental decoherence-dissipation. We studied the area under the concurrence graph of the first-largest cycle of entanglement, in order to gain information for both the magnitude of entanglement production  and the disentanglement time. Indeed, there is no unique measure of stochastic resonance-like effects and one has to decide which is the best for the physical property under study.

Moreover, temperature can play a constructive role for the control of the special initial Hilbert subspaces which lead to non-monotonic noise effects. It provides us with an easy way to cut out the first-part of S.AR..

Of course the number of different preparations we tried is not large, but it is enough to show the main results, which strengthen our belief that:
\begin{enumerate}
\item{Noise effects are not fully understood yet, even in the Markovian regime which simplifies significantly the mathematical calculations.}
\item{Initial system preparation (and hence the local geometry of the state space) may allow noise to affect decoherence in a positive or negative way.}
\item{The initial local state space of qubit systems needs to be mapped according to the noise effects it can exhibit. The classification of these subspaces  could lead to  predictions of non-mo\-no\-tonic behavior, and consequently to effective optimization of a quantum computation process.}
\end{enumerate}

\appendix
\section{3-qubit chain density matrix and summary tables}
The general form of the 3-qubit density matrix is:

\begin{eqnarray}\label{d3}
\rho_{s}=\left[ \begin{array}{*{20}{c}}
 \bullet &&& \bullet && \bullet & \bullet &\\
& \bullet & \bullet && \bullet &&& \bullet \\
& \bullet & \bullet && \bullet &&& \bullet \\
 \bullet &&& \bullet && \bullet & \bullet &\\
& \bullet & \bullet && \bullet &&& \bullet \\
 \bullet &&& \bullet && \bullet & \bullet &\\
 \bullet &&& \bullet && \bullet & \bullet &\\
& \bullet & \bullet && \bullet &&& \bullet
\end{array} \right],
\end{eqnarray}
where all blank elements are equal to zero.

Next we present summary tables for the 3-qubit case, for several initial preparations of product states and Bell states.

\begin{table}[ht8]
\caption{\label{a3q1} Product state results}
\begin{indented}
\item[]\begin{tabular}{@{} l l l l l}
\br
Initial State & N. S. & S. AR. & S. R. & Noise \\
\mr
$|eee\rangle\to |ee\rangle_{12}$ &  &  &  &  \\\mr
$|eeg\rangle\to |ee\rangle_{12}$ & \checkmark &  & & $M_3$ \\\mr
$|ege\rangle\to |eg\rangle_{12}$ &  &\checkmark & &  $M_3$\\\mr
$|egg\rangle\to |eg\rangle_{12}$ &  &\checkmark &\checkmark & $M_3$\\\mr
$|gee\rangle\to |ge\rangle_{12}$ &  &\checkmark& & $M_3$\\\mr
$|geg\rangle\to |ge\rangle_{12}$ &  & \checkmark &\checkmark & $M_3$ \\\mr
$|gge\rangle\to |gg\rangle_{12}$ & \checkmark &  &  &  $M_3$\\\mr
$|ggg\rangle\to |gg\rangle_{12}$ &  &  & & \\
\br
\end{tabular}
\end{indented}
\end{table}
\clearpage

\begin{table}[ht7]
\caption{\label{a3q2} Bell state $|\Phi^+\rangle=\frac{|ee\rangle+|gg\rangle}{\sqrt{2}}$}
\begin{indented}
\item[]\begin{tabular}{@{} l l l l l}
\br
Initial State & N. S. & S. AR. & S. R. & Noise \\
\mr
$\rho_{1,1}=\rho_{2,2}=\rho_{7,7}=\rho_{8,8}=1/4$& \checkmark & &  & $M_3$ \\
$\rho_{1,7}=\rho_{7,1}=\rho_{2,8}=\rho_{8,2}=1/4$&&&&\\\mr
$\rho_{1,1}=0,\rho_{2,2}=1/2,\rho_{7,7}=0,\rho_{8,8}=1/2$& \checkmark & &  & $M_3$ \\
$\rho_{1,7}=\rho_{7,1}=0,\rho_{2,8}=\rho_{8,2}=1/2$&&&&\\\mr
$\rho_{1,1}=2/5,\rho_{2,2}=1/10,\rho_{7,7}=2/5,\rho_{8,8}=1/10$& \checkmark & &  & $M_3$ \\
$\rho_{1,7}=\rho_{7,1}= 2/5,\rho_{2,8}=\rho_{8,2}=1/10$&&&&\\\mr
$\rho_{1,1}=1/10,\rho_{2,2}=2/5,\rho_{7,7}=1/10,\rho_{8,8}=2/5$& \checkmark & &  & $M_3$ \\
$\rho_{1,7}=\rho_{7,1}= 1/10,\rho_{2,8}=\rho_{8,2}=2/5$&&&&\\\mr
$\rho_{1,1}=1/2,\rho_{2,2}=0,\rho_{7,7}=1/2,\rho_{8,8}=0$& \checkmark & &  & $M_3$ \\
$\rho_{1,7}=\rho_{7,1}= 1/2,\rho_{2,8}=\rho_{8,2}=0$&&&&\\
\br
\end{tabular}
\end{indented}
\vspace{1cm}
\caption{\label{a3q3} Bell state $|\Psi^+\rangle=\frac{|eg\rangle+|ge\rangle}{\sqrt{2}}$}
\begin{indented}
\item[]\begin{tabular}{@{} l l l l l}
\br
Initial State & N. S. & S. AR. & S. R. & Noise \\
\mr
$\rho_{3,3}=\rho_{4,4}=\rho_{5,5}=\rho_{6,6}=1/4$& \checkmark & &  & $M_3$ \\
$\rho_{3,5}=\rho_{5,3}=\rho_{4,6}=\rho_{6,4}=1/4$&&&&\\\mr
$\rho_{3,3}=0,\rho_{4,4}=1/2,\rho_{5,5}=0,\rho_{6,6}=1/2$&  & & \checkmark & $M_3$ \\
$\rho_{3,5}=\rho_{5,3}=0,\rho_{4,6}=\rho_{6,4}=1/2$&&&&\\\mr
$\rho_{3,3}=1/2,\rho_{4,4}=0,\rho_{5,5}=1/2,\rho_{6,6}=0$& \checkmark & &  & $M_3$ \\
$\rho_{3,5}=\rho_{5,3}=1/2,\rho_{4,6}=\rho_{6,4}=0$&&&&\\\mr
$\rho_{3,3}=2/5,\rho_{4,4}=1/10,\rho_{5,5}=2/5,\rho_{6,6}=1/10$& \checkmark & &  & $M_3$ \\
$\rho_{3,5}=\rho_{5,3}=2/5,\rho_{4,6}=\rho_{6,4}=1/10$&&&&\\\mr
$\rho_{3,3}=1/10,\rho_{4,4}=2/5,\rho_{5,5}=1/10,\rho_{6,6}=2/5$& \checkmark & &  & $M_3$ \\
$\rho_{3,5}=\rho_{5,3}=1/10,\rho_{4,6}=\rho_{6,4}=2/5$&&&&\\
\br
\end{tabular}
\end{indented}
\end{table}

\clearpage
\section{4-qubit chain density matrix and summary tables}
The general form of the 4-qubit density matrix is:

\begin{eqnarray}\label{d4}
\rho_{s}=
\left[ \begin{array}{*{20}{c}}
 \bullet &&& \bullet && \bullet & \bullet &&& \bullet & \bullet && \bullet &&& \bullet \\
& \bullet & \bullet && \bullet &&& \bullet & \bullet &&& \bullet && \bullet & \bullet &\\
& \bullet & \bullet && \bullet &&& \bullet & \bullet &&& \bullet && \bullet & \bullet &\\
 \bullet &&& \bullet && \bullet & \bullet &&& \bullet & \bullet && \bullet &&& \bullet \\
& \bullet & \bullet && \bullet &&& \bullet & \bullet &&& \bullet && \bullet & \bullet &\\
 \bullet &&& \bullet && \bullet & \bullet &&& \bullet & \bullet && \bullet &&& \bullet \\
 \bullet &&& \bullet && \bullet & \bullet &&& \bullet & \bullet && \bullet &&& \bullet \\
& \bullet & \bullet && \bullet &&& \bullet & \bullet &&& \bullet && \bullet & \bullet &\\
& \bullet & \bullet && \bullet &&& \bullet & \bullet &&& \bullet && \bullet & \bullet &\\
 \bullet &&& \bullet && \bullet & \bullet &&& \bullet & \bullet && \bullet &&& \bullet \\
 \bullet &&& \bullet && \bullet & \bullet &&& \bullet & \bullet && \bullet &&& \bullet \\
& \bullet & \bullet && \bullet &&& \bullet & \bullet &&& \bullet && \bullet & \bullet &\\
 \bullet &&& \bullet && \bullet & \bullet &&& \bullet & \bullet && \bullet &&& \bullet \\
& \bullet & \bullet && \bullet &&& \bullet & \bullet &&& \bullet && \bullet & \bullet &\\
& \bullet & \bullet && \bullet &&& \bullet & \bullet &&& \bullet && \bullet & \bullet &\\
 \bullet &&& \bullet && \bullet & \bullet &&& \bullet & \bullet && \bullet &&& \bullet
\end{array} \right],
\end{eqnarray}
where all blank elements are equal to zero.

Next we present summary tables for the 4-qubit case, for several initial preparations of product states and Bell states.
\clearpage
\begin{table}[ht4]
\caption{\label{a4q1} Product states}
\begin{indented}
\item\begin{tabular}{@{} l l l l l p{6cm}}
\br
Initial State & N. S. & S. AR. & S. R. & Noise \\
\mr
$|eeee\rangle\to |ee\rangle_{12}$ &   &  &  \\
\mr
$|eeeg\rangle\to |ee\rangle_{12}$ & \checkmark &  &  &$M_{34},M_4$  \\
\mr
$|eege\rangle\to |ee\rangle_{12}$ & \checkmark & &  & $M_{34}$  \\
\mr
$|eegg\rangle\to |ee\rangle_{12}$ & \checkmark & &  & $M_{34},M_4$ \\
\mr
$|egee\rangle\to |ee\rangle_{12}$ & \checkmark & &  & $M_{34},M_4$  \\
\mr
$|egeg\rangle\to |eg\rangle_{12}$ & &\checkmark &  & $M_{34},M_4$  \\
\mr
$|egge\rangle\to |eg\rangle_{12}$ &  \checkmark & &  & $M_{34},M_4$  \\
\mr
$|eggg\rangle\to |eg\rangle_{12}$ &  \checkmark & &  & $M_{34},M_4$  \\
\mr
$|geee\rangle\to |ge\rangle_{12}$ &  \checkmark & &  & $M_{34},M_4$  \\
\mr
$|geeg\rangle\to |ge\rangle_{12}$ &  \checkmark & &  & $M_{34},M_4$  \\
\mr
$|gege\rangle\to |ge\rangle_{12}$ & \checkmark &\checkmark & & N.S.$\to M_4$, S.AR. $\to M_{34}$  \\
\mr
$|gegg\rangle\to |ge\rangle_{12}$ & \checkmark & \checkmark & \checkmark &  N.S $\to$ $M_{34}$,  S.AR. + S.R. $\to$ $M_{4}$ \\
\mr
$|ggee\rangle\to |gg\rangle_{12}$  & \checkmark & & & $M_{34}$, $M_{4}$  \\
\mr
$|ggeg\rangle\to |gg\rangle_{12}$ &  &\checkmark  & \checkmark & S.AR. + S.R. $\to$ $M_{1234}$, \\
& & & &  $M_{234}$, $M_{12}$, $M_{13}$,\\& & & &  $M_{14}$, $M_1$\\
\mr
$|ggge\rangle\to |gg\rangle_{12}$ &\checkmark   & \checkmark  & \checkmark & N.S $\to$ $M_{4}$\\
& & & & S.AR. + S.R. $\to$ $M_{1234}$, \\
& & & &  $M_{234}$, or $M_{12}$, $M_{13}$,\\& & & & $M_{14}$, $M_{1}$\\
\mr
$|gggg\rangle\to |gg\rangle_{12}$ &  & & & \\
\br
\end{tabular}
\end{indented}
\end{table}

S.AR. + S.R. means that there exist resonance and anti-resonance in the same diagram (i.e. \fref{3q}(a)).
\clearpage
\begin{table}[ht5]
\caption{\label{a4q2} Bell state $|\Phi^+\rangle=\frac{|ee\rangle+|gg\rangle}{\sqrt{2}}$}
\begin{indented}
\item[]\begin{tabular}{@{} l l l l l}
\br
Initial State & N. S. & S. AR. & S. R. & Noise \\
\mr
$\rho_{1,1}=\rho_{2,2}=\rho_{3,3}=\rho_{4,4}=\frac{1}{8}$& &&& \\
$\rho_{13,13}=\rho_{14,14}=\rho_{15,15}=\rho_{16,16}=\frac{1}{8}$&\checkmark & &   &$M_{34}$, $M_4$\\
$\rho_{13,1}=\rho_{14,2}=\rho_{15,3}=\rho_{16,4}=\frac{1}{8}$&&&&\\
$\rho_{1,13}=\rho_{2,14}=\rho_{3,15}=\rho_{4,16}=\frac{1}{8}$&&&&\\\mr
$\rho_{1,1}=\frac{1}{2},\rho_{2,2}=\rho_{3,3}=\rho_{4,4}=0$& &&& \\
$\rho_{13,13}=\frac{1}{2},\rho_{14,14}=\rho_{15,15}=\rho_{16,16}=0$&  &\checkmark &   &$M_{34}$ \\
$\rho_{13,1}=\rho_{1,13}=\frac{1}{2},\rho_{14,2}=\rho_{15,3}=0$&&&&\\
$\rho_{2,14}=\rho_{3,15}=\rho_{4,16}=\rho_{16,4}=0$&&&&\\\mr
$\rho_{1,1}=0,\rho_{2,2}=\frac{1}{2},\rho_{3,3}=\rho_{4,4}=0$& &&& \\
$\rho_{14,14}=\frac{1}{2},\rho_{13,13}=\rho_{15,15}=\rho_{16,16}=0$&\checkmark &  &  &$M_{34}$, $M_4$ \\
$\rho_{13,1}=\rho_{1,13}=0,\rho_{2,14}=\rho_{14,2}=\frac{1}{2},$&&&&\\
$\rho_{3,15}=\rho_{15,3}=\rho_{4,16}=\rho_{16,4}=0$&&&&\\\mr
$\rho_{1,1}=\rho_{2,2}=0,\rho_{3,3}=\rho_{4,4}=\frac{1}{4}$& &&& \\
$\rho_{13,13}=\rho_{14,14}=0,\rho_{15,15}=\rho_{16,16}=\frac{1}{4}$&\checkmark &  &   &$M_{34}$, $M_4$ \\
$\rho_{13,1}=\rho_{1,13}=\rho_{2,14}=\rho_{14,2}=0,$&&&&\\
$\rho_{3,15}=\rho_{15,3}=\rho_{4,16}=\rho_{16,4}=\frac{1}{4}$&&&&\\\mr
$\rho_{1,1}=\rho_{2,2}=\frac{1}{4},\rho_{33}=\rho_{4,4}=0$& &&& \\
$\rho_{13,13}=\rho_{14,14}=\frac{1}{4},\rho_{15,15}=\rho_{16,16}=0$&\checkmark & &   &$M_{34}$, $M_4$ \\
$\rho_{13,1}=\rho_{1,13}=\rho_{2,14}=\rho_{14,2}=\frac{1}{4}$&&&&\\
$\rho_{3,15}=\rho_{15,3}=\rho_{4,16}=\rho_{16,4}=0$&&&&\\\mr
$\rho_{1,1}=\rho_{2,2}=\rho_{3,3}=0,\rho_{4,4}=\frac{1}{2}$& &&& \\
$\rho_{13,13}=\rho_{14,14}=\rho_{15,15}=0,\rho_{16,16}=\frac{1}{2}$& &\checkmark &  &  $ M_{34}$ \\
$\rho_{13,1}=\rho_{1,13}=\rho_{2,14}=\rho_{14,2}=0$&&&&\\
$\rho_{3,15}=\rho_{15,3}=0,\rho_{4,16}=\rho_{16,4}=\frac{1}{2}$&&&&\\\mr
$\rho_{1,1}=\rho_{2,2}=\frac{1}{16},\rho_{3,3}=\rho_{4,4}=\frac{3}{16}$& &&& \\
$\rho_{13,13}=\rho_{14,14}=\frac{1}{16},\rho_{15,15}=\rho_{16,16}=\frac{3}{16}$&\checkmark & &  &  $ M_{34},M_4$ \\
$\rho_{13,1}=\rho_{1,13}=\rho_{2,14}=\rho_{14,2}=\frac{1}{16}$&&&&\\
$\rho_{3,15}=\rho_{15,3}=\rho_{4,16}=\rho_{16,4}=\frac{3}{16}$&&&&\\
\br
\end{tabular}
\end{indented}
\end{table}

\clearpage

\begin{table}[ht]
\caption{ \label{a4q3}Bell state $|\Psi^+\rangle=\frac{|eg\rangle+|ge\rangle}{\sqrt{2}}$}
\begin{tabular}{@{} l l l l l}
\br
Initial State & N.S. & S.AR. & S.R. & Noise \\
\mr
$\rho_{5,5}=\rho_{6,6}=\rho_{7,7}=\rho_{8,8}=\frac{1}{8}$& &&& \\
$\rho_{9,9}=\rho_{10,10}=\rho_{11,11}=\rho_{12,12}=\frac{1}{8}$&\checkmark & &   &$M_{34},M_4$\\
$\rho_{5,9}=\rho_{6,10}=\rho_{7,11}=\rho_{8,12}=\frac{1}{8}$&&&&\\
$\rho_{9,5}=\rho_{10,6}=\rho_{11,7}=\rho_{12,8}=\frac{1}{8}$&&&&\\\mr
$\rho_{5,5}=\rho_{6,6}=\frac{3}{16},\rho_{7,7}=\rho_{8,8}=\frac{1}{16}$& &&& \\
$\rho_{9,9}=\rho_{10,10}=\frac{3}{16},\rho_{11,11}=\rho_{12,12}=\frac{1}{16}$&\checkmark & &   &$M_{34},M_4$\\
$\rho_{5,9}=\rho_{6,10}=\frac{3}{16},\rho_{7,11}=\rho_{8,12}=\frac{1}{16}$&&&&\\
$\rho_{9,5}=\rho_{10,6}=\frac{3}{16},\rho_{11,7}=\rho_{12,8}=\frac{1}{16}$&&&&\\\mr
$\rho_{5,5}=\rho_{6,6}=\frac{1}{16},\rho_{7,7}=\rho_{8,8}=\frac{3}{16}$& &&& \\
$\rho_{9,9}=\rho_{10,10}=\frac{1}{16},\rho_{11,11}=\rho_{12,12}=\frac{3}{16}$&\checkmark & &   &$M_{34},M_4$\\
$\rho_{5,9}=\rho_{6,10}=\frac{1}{16},\rho_{7,11}=\rho_{8,12}=\frac{3}{16}$&&&&\\
$\rho_{9,5}=\rho_{10,6}=\frac{1}{16},\rho_{11,7}=\rho_{12,8}=\frac{3}{16}$&&&&\\\mr
$\rho_{5,5}=\frac{1}{2},\rho_{6,6}=\rho_{7,7}=\rho_{8,8}=0$& &&& \\
$\rho_{9,9}=\frac{1}{2},\rho_{10,10}=\rho_{11,11}=\rho_{12,12}=0$&\checkmark & &   &$M_{34},M_4$\\
$\rho_{5,9}=\frac{1}{2},\rho_{6,10}=\rho_{7,11}=\rho_{8,12}=0$&&&&\\
$\rho_{9,5}=\frac{1}{2},\rho_{10,6}=\rho_{11,7}=\rho_{12,8}=0$&&&&\\\mr
$\rho_{5,5}=0,\rho_{6,6}=\frac{1}{2},\rho_{7,7}=\rho_{8,8}=0$& &&& \\
$\rho_{9,9}=0,\rho_{10,10}=\frac{1}{2},\rho_{11,11}=\rho_{12,12}=0$& &\checkmark &   &$M_{34},M_4$ \\
$\rho_{5,9}=0,\rho_{6,10}=\frac{1}{2},\rho_{7,11}=\rho_{8,12}=0$&&&&\\
$\rho_{9,5}=0,\rho_{10,6}=\frac{1}{2},\rho_{11,7}=\rho_{12,8}=0$&&&&\\\mr
$\rho_{5,5}=\rho_{6,6}=0,\rho_{7,7}=\frac{1}{2},\rho_{8,8}=0$& &&& \\
$\rho_{9,9}=\rho_{10,10}=0,\rho_{11,11}=\frac{1}{2},\rho_{12,12}=0$& &\checkmark & \checkmark  &S.AR. $\to M_{34}$ \\&&&&S.R.+S.AR. $\to M_4$ \\
$\rho_{5,9}=\rho_{6,10}=0,\rho_{7,11}=\frac{1}{2},\rho_{8,12}=0$&&&&\\
$\rho_{9,5}=\rho_{10,6}=0,\rho_{11,7}=\frac{1}{2},\rho_{12,8}=0$&&&&\\\mr
$\rho_{5,5}=\rho_{6,6}=\rho_{7,7}=0,\rho_{8,8}=\frac{1}{2}$ \\
$\rho_{9,9}=\rho_{10,10}=\rho_{11,11}=0,\rho_{12,12}=\frac{1}{2}$&\checkmark & & \checkmark  &\parbox[t]{4cm}{N.S $\to M_{34},M_4$\\ S.R. $\to M_{234},M_2$} \\
$\rho_{5,9}=\rho_{6,10}=\rho_{7,11}=0,\rho_{8,12}=\frac{1}{2}$\\
$\rho_{9,5}=\rho_{10,6}=\rho_{11,7}=0,\rho_{12,8}=\frac{1}{2}$\\
\br
\end{tabular}
\end{table}

\end{document}